\documentclass[aps,prb,twocolumn,superscriptaddress]{revtex4-1}
\usepackage{graphicx}
\usepackage{bm}
\usepackage{amsmath}
\usepackage{amssymb}
\usepackage{euscript}
\usepackage{booktabs,array}
\usepackage{verbatim}
\usepackage{fancyhdr}
\usepackage{wrapfig}
\usepackage{setspace}
\usepackage{xcolor}
\usepackage{amsfonts}
\usepackage{subfigure}
\usepackage{nicefrac}
\usepackage{array,booktabs}
\usepackage{hhline}
\newcommand{\be}{\begin{equation}}
\newcommand{\ee}{\end{equation}}
\newcommand{\bea}{\begin{eqnarray}}
\newcommand{\eea}{\end{eqnarray}}

\begin{document}

	\title{SO(5) critical point in a spin-flavor Kondo device - Bosonization and refermionization solution}

	\author{Alon Liberman}
	\affiliation{School of Physics and Astronomy,
		Tel Aviv University, Tel Aviv 6997801, Israel}
	
	\author{Andrew K. Mitchell}
	\affiliation{School of Physics, University College Dublin, Belfield, Dublin 4, Ireland} 
	\author{Ian Affleck}
	\affiliation{Department of Physics and Astronomy and Stewart Blusson Quantum Matter Institute, 
		University of British Columbia, Vancouver, B.C., Canada, V6T 1Z1}

	\author{Eran Sela}
	\affiliation{School of Physics and Astronomy,
		Tel Aviv University, Tel Aviv 6997801, Israel}
	
	\begin{abstract}
		We investigate a well studied system of a quantum dot coupled to a Coulomb box and leads, realizing a spin-flavor Kondo model. It exhibits a recently discovered non-Fermi liquid (NFL) behavior with emergent SO(5) symmetry~\cite{mitchell2020so}. 
		Here, through a detailed bosonization and refermionization solution,
		we push forward our previous work and provide a  consistent and complete description of the various exotic properties and phase diagram. A unique NFL phase emerges from the presence of an uncoupled Majorana fermion from the flavor sector, whereas FL-like susceptibilities result from the gapping out of a pair of Majroana fermions from the spin and flavor sectors. Other properties, such as a $T^{3/2}$ scaling of the conductance, stability under channel or spin symmetry breaking and a re-appearance of NFL behavior upon breaking the particle-hole symmetry, are all accounted for by a renormalization group treatment of the  refermionized Majorana model.      
	\end{abstract}
	
	\maketitle
	
	\linespread{1}

	\section{Introduction}
	Quantum dots (QDs) are among the most basic building blocks of mesoscopic circuits~\cite{sohn2013mesoscopic}, providing a test-bed for strongly correlated and entangled problems within the simple context of a quantum impurity models.
	The physical properties of QDs depend essentially
	on their level spacing $\delta$, charging energy $E_c$, and precise form of the coupling to
	their surroundings: They can exhibit Coulomb blockade phenomena at low temperatures~\cite{wilkins1989scanning} $T \ll E_c$ and build up entangled Kondo-like states of various
	kinds~\cite{goldhaber1998kondo,van2000kondo,cronenwett1998tunable,cronenwett2002low,glazman1990lifting} below the Kondo temperature $T \ll T_K$.
	Within the Coulomb blockade regime, one may either have small QDs dominated by a single quantum level due to a large level spacing $\delta \gg T$, or large QDs which are metallic grains with $\delta \ll T$, referred to here as ``Coulomb boxes", having a large density of states, yet displaying charge quantization~\cite{matveev1995coulomb,furusaki1995theory}. 
	
	This combination of large charging energy and large density of states has been a the central ingredient in the first experimental realization of the two-channel Kondo (2CK) effect~\cite{oreg2003two,potok2007observation,keller2015universal}.  Here, multiple Coulomb boxes act as effective screening channels of a small central QD carrying an unpaired spin, and the boxes' charging energy completely suppresses inter-channel charge transfer. Its exotic non-Fermi liquid (NFL) behavior is reflected in non-trivial electronic scattering properties which were calculated using conformal field theory (CFT)~\cite{affleck1993exact} accounting for anomalous experimental signatures~\cite{potok2007observation,keller2015universal}.

	These experiments opened a line of research activity on the quantum critical nature of this NFL state~\cite{pustilnik2004quantum,vojta2006impurity,toth2007dynamical,toth2008dynamical,sela2011exact}, the role of charge fluctuations near charge degeneracy of the Coulomb box~\cite{le2003smearing,le2004maximized,anders2005coulomb}, the non-local role of the Coulomb charging energy~\cite{florens2004climbing}, various transport properties~\cite{simon2006transport,liu2008transport,carmi2012transmission,mitchell2012universal}, capacitance signatures~\cite{bolech2005prediction} multiple impurities generalizations~\cite{mitchell2011two} and related devices~\cite{kikoin2007magnetic}. More recently, Coulomb boxes were implemented in the strong magnetic field regime~\cite{iftikhar2015two,iftikhar2018tunable,anthore2018circuit} leading to a convenient  experimental platform to study multichannel charge-Kondo effects achieved near charge degeneracy points of the Coulomb box~\cite{furusaki1995theory,mitchell2016universality,landau2018charge,iftikhar2018tunable,van2019wiedemann,nguyen2020thermoelectric,lee2020fractional}.
	
	The lead-dot-box device of Refs.~\onlinecite{oreg2003two,potok2007observation,keller2015universal} on which we focus, see Fig.~\ref{fig:phase_diagram}, exhibits a rich phase diagram invoking correlations between spin and charge degrees of freedom, as a function of gate parameters controlling the charges of the box and small QD. In particular, a certain tuning of parameters gives rise to a spin-flavor Kondo effect: a phase in which the spin-1/2 formed in the small QD, $\vec{S}$, gets entangled with a ``flavor" pseudospin-1/2 operator $\vec{T}$, associated with two charge states of the box.  
	
	To model the lead-dot-box system in Fig.~\ref{fig:phase_diagram} we study the low energy Hamiltonian $H_0+\delta H$, where $H_0=\sum_k \sum_{\alpha=L,B,\sigma=\uparrow,\downarrow} \epsilon_k c^{\dagger}_{k \alpha \sigma } c_{k \alpha \sigma}$ describes the leads ($\alpha=L$) or the box ($\alpha=B$), respectively, and 
	\bea
	\label{hamiltonianlehur}
	\delta H&=& \sum_{a=x,y,z} J \psi^\dagger \frac{\sigma^a}{2}  \psi S^a +\sum_{b=x,y,z}  V_b \psi^\dagger \frac{\tau^b}{2}  \psi T^b  \nonumber \\
	&+&\sum_{a,b=x,y,z}  Q_b \psi^\dagger \sigma^a \tau^b  \psi S^a T^b, \nonumber \\
	&&Q_x=Q_y \equiv Q_\perp,~~~V_x=V_y \equiv V_\perp.
	\eea
	Here Pauli matrices $\sigma^a$ and $\tau^b$ act in the spin and flavor sectors, respectively. In Eq.~(\ref{hamiltonianlehur}) we suppress  spin and flavor indices,  for example $\psi^\dagger \sigma^a \tau^b \psi \equiv   \sum_{\alpha,\beta, \sigma ,\sigma' } \psi^{\dagger}_{ \alpha \sigma } \sigma^a_{\sigma \sigma'} \tau^b_{\alpha \beta}\psi_{ \beta \sigma'}$ where we define local field operators $\psi_{\alpha \sigma}=\sum_k c_{k \alpha \sigma}$. 
	In this paper, we study the rich phase diagram of this Hamiltonian with a number of additional perturbations.

	\subsection{Previous results and emergent symmetry}
	An earlier work on this model by Borda \emph{et. al.}~\cite{borda20034} found that at low energies, the ratios between the various coupling constants $J, Q_b, V_b$ ($b=x,y,z=1,2,3$) flow under renormalization group (RG) to unity. Remarkably, the resulting low temperature fixed point has an enlarged SU(4) symmetry. Using numerical RG (NRG) and CFT it was found that this SU(4) symmetric state is a stable FL. This high symmetry state featured in a number of QD experiments~\cite{jarillo2005orbital,makarovski20084}. 
	
	\begin{figure}
		\includegraphics[width=6cm]{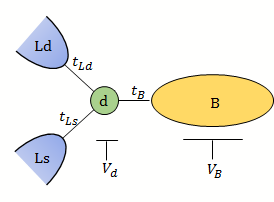}
		\caption{A quantum dot ($d$) coupled to a quantum box (B) and source/drain leads (Ls/Ld). }
		\label{fig:phase_diagram}
	\end{figure}

	Le Hur \emph{et. al}~\cite{le2003smearing,le2004maximized} explored the consequences of this stable SU(4) FL state on the conductance and capacitance.
	The model in Fig.~\ref{fig:phase_diagram} was also  elaborated on in a series of papers by Anders, Lebanon and Schiller (ALS)~\cite{lebanon2003enhancement,lebanon2003coulomb,anders2004coulomb}. They focused on the particle-hole (PH) symmetric case and found a NFL ground state, rather than a FL behavior claimed by Le Hur \emph{et. al.}. Also, ALS found a smooth crossover from spin- to flavor 2CK NFL behavior as a function of the box's gate voltage. 
	
	Below,  
	these two seemingly conflicting results will be reconciled by the introduction of a novel NFL fixed point, located at the PH symmetric point and  characterized by a SO(5) symmetry~\cite{mitchell2020so}, exhibiting both FL-like susceptibilities and NFL fractional entropy, along with a $T^{\frac{3}{2}}$ scaling of the conductance. While the key features of this SO(5) fixed point were found in Ref.~\onlinecite{mitchell2020so}, here we provide a detailed analysis of the phase diagram and interplay of different perturbations.

	Our main endeavor  is the investigation of this unique SO(5) point along with it's stability to the following perturbations: magnetic field ($\Delta H_{B}=BS^{z}$), flavor field ($\Delta H_{B_{f}}=B_{f}T^{z}$),  channel symmetry breaking  ($\Delta H_{as}=J_{-}\psi^{\dagger}\frac{\vec{\sigma}\tau^{z}}{2}\psi\vec{S}$) and PH symmetry breaking  [nonzero $V_{\perp}$ and $Q_{z}$ in Eq.~\eqref{hamiltonianlehur}].
	We show that the NFL phase at the SO(5) point is stable to the inclusion of a magnetic field or channel symmetry breaking. While this stability of the NFL state is reflected by a fractional entropy of $\frac{1}{2} \log 2$, other quantities such as the spin susceptibility show FL-like behavior~\cite{le2003smearing,le2004maximized} of an inherently NFL state. 
	
\begin{figure}
		\includegraphics[width=7cm]{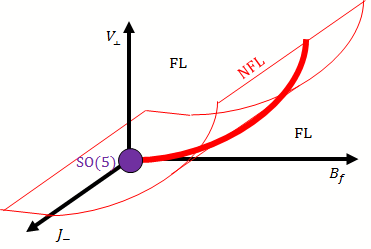}
		\caption{(Color online) A schematic drawing of the NFL line  (in bold) emanating from the SO(5) fixed point   in the plane spanned by $B_{f}$ and $V_{\perp}$ perturbations. The NFL line becomes a NFL 2D manifold as we add the $J_{-}$ term as a third perturbation. We describe this curve explicitly in Sec.~VI using Eq.~\eqref{eq:V0}.}
		\label{phase}
	\end{figure}

	The addition of PH breaking including a flavor field, however, can destabilize the NFL to a FL. Interestingly, when both destabilizing perturbations occur, it is still possible to tune them to cancel each other and re-stabilize a NFL phase, which persists along a curve in the plane spanned by the PH breaking perturbation and flavor field perturbation [bold curve in Fig.~\ref{phase}]. In the QD device in Fig.~1 this corresponds to a curve in the phase diagram spanned by the gate voltages $V_d$ and $V_B$ [drawn in Fig.~\ref{gf:figfit} for different parameter choices]. Adding a third axis representing e.g. $\Delta H_{as}$, results in a 2D NFL manifold in Fig.~\ref{phase}. This means that the NFL line in the phase diagram exists for generic ratios between the tunneling coefficients of the leads and the box~\cite{mitchell2020so}. 
	
The plan of the paper is as follows. In Sec.~II we present the lead-dot-box Hamiltonian and map it to the low energy form of Eq.~\eqref{hamiltonianlehur} by the Schrieffer-Wolff transformation. We  examine the system's RG behavior with and without  PH breaking terms, and make the emergent symmetry explicit. 
In Sec.~III we apply bosonization and refermionization techniques, obtaining a Majorana representation of the problem.
In Sec.~IV, we analyze the spin-flavor phase at the SO(5) point in terms of thermodynamic quantities (entropy, spin and flavor susceptibilities) and conductance. In Sec.~V, we explore how these quantities change under different perturbations, and in Sec. VI we combine PH symmetry breaking perturbations to show the emergence of a NFL line from the SO(5) point in the phase diagram. Our field theory results in Sections.~V and VI are compared with our NRG calculations. We briefly conclude in Sec.~VII.

\section{Model}
\subsection{Lead-dot-box model}
As shown in Fig.~\ref{fig:phase_diagram}, our system, studied in numerous earlier works~\cite{oreg2003two,pustilnik2004quantum,le2004maximized,anders2005coulomb}, consists of a central QD ($d$)  connected to source and drain leads ($Ls$,~$Ld$), and to a quantum box ($B$). It is described by the Hamiltonian $H=H_0+H_{ B}+H_{ d}+H_{hyb}$. The  hybridization term $H_{hyb}=\sum_{\alpha}H_{hyb}^{\alpha}$ couples the small dot  with the various reservoirs. Here $H_0=\sum_{k,\alpha,\sigma}\epsilon_{k}^{\phantom{\dagger}}  c_{k \alpha \sigma}^{\dagger}c_{k \alpha \sigma}^{\phantom{\dagger}}$ describes the three conduction electron reservoirs,   
\begin{align}
H_{ B} &= E_c \left ( \hat{N}_{ B} - N_0 - n_g \right )^2 \;, \label{eq:Hb} \\
H_{ d} &=  \epsilon_{d} n_d+ U  n_\uparrow n_\downarrow   \;, \label{eq:Hd}
\end{align}
describe the box Coulomb interaction and the Hamiltonian of the small dot, respectively, and the hybridization term is given by $H_{\rm hyb}^{\alpha}=\sum_{k,\sigma}(t_{k \alpha }d_{\sigma}^{\dagger}c_{k \alpha \sigma}^{\phantom{\dagger}} +   H.c.)$. Here, $\sigma =\uparrow,\downarrow$ denotes (real) spin, and $d_{\sigma}$ or $c_{k \alpha \sigma}$ are annihilation operators for the dot or conduction electrons, respectively. $n_\sigma=d^\dagger_\sigma d_\sigma$ and $n_d=\sum_\sigma n_\sigma$ measure the number of electrons in the dot, and $\hat{N}_{ B}=\sum_{k,\sigma} c_{k B  \sigma}^{\dagger}c_{k B \sigma}^{\phantom{\dagger}}$ is the  number operator for the box. 

The dot and box occupations are controllable by gate voltages
\be
V_{d} \propto \eta =\epsilon_{ d}+\tfrac{1}{2}U,~~~ 
V_{ B} \propto n_{ g}-\frac{1}{2},
\ee 
respectively, see Fig.~1.  
We take equivalent conduction electron baths $\epsilon_{\alpha k}\equiv \epsilon_k$ with a constant density of states $\nu$ (which we set to unity).
We define $t_{\alpha}^2 = \sum_k |t_{\alpha k}|^2$ and $t_{ L}^2=t_{ Ls}^2+ t_{Ld}^2$. The two leads can be effectively treated as a single lead. 
Thus, the combined lead ($L$) and box ($B$) act as two channels~\cite{pustilnik2004quantum}. We will often refer to the lead and box as the left ($L$) and right ($R$) channels.
At $n_{g}=\tfrac{1}{2}$, the box states with $N_0$ and $N_0+1$ electrons are  degenerate. Neglecting other high-energy box charge states 
we may define a pseudospin-$\tfrac{1}{2}$ operator $T^{+}=|N_0+1\rangle\langle N_0|$, $T^{-}=(T^+)^{\dagger}$, and $T^z=\tfrac{1}{2}(|N_0+1\rangle\langle N_0+1| - |N_0\rangle\langle N_0|)$. The charge pseudospin is flipped by electronic tunneling between the dot and box. The lead-dot-box Hamiltonian is then
\bea
\label{eq:HALS1}
H_{ALS}&=& \sum_{\alpha = L,R} \sum_{k, \sigma} \epsilon_{k} c^\dagger_{k \alpha   \sigma} c_{k \alpha  \sigma}+  E_c ( T^z+1/2-n_{g})^2 \nonumber  \\ 
&+&H_{d}+\sum_{k , \sigma} [d_\sigma^\dagger (t_{L} c_{k L  \sigma}+t_{B} c_{k B  \sigma} T^- ) +H.c.]. 
\eea
We refer to this model in which the tunneling terms are supplemented by the $T-$pseudo-spin operator as the Anders, Lebanon and Schiller (ALS) model. 

\subsection{PH transformation}
In the next subsection we map the lead-dot-box model to the spin-flavor Kondo model, Eq.~\eqref{hamiltonianlehur}, perturbed by various terms. Before doing so, we introduce a PH transformation allowing us to distinguish  various interactions. The PH transformation takes
\bea
\label{ourPH}
\psi & \to &  \sigma^y\psi^\dagger,~~~~~~   d  \to  -\sigma^y d^\dagger, \\
T^+  & \to & T^- ,~~~~~ T^z  \to - T^z . \nonumber
\eea
For example, this symmetry takes $\psi^\dagger  \tau^z\psi
\to -\psi^\dagger \tau^z\psi$ so the $(\psi^\dagger  \tau^z \psi) T^z$ term (associated with the $V_{z}$ coupling) respects the symmetry since 
$T^z$
also changes sign. On the other hand, it takes $\psi^\dagger \tau^+\psi
\to -\psi^\dagger \tau^-\psi$ and $T^+\to T^-$ so the $\pm$-flavor Kondo
terms (with coupling $V_{\perp}$) change sign and are PH odd. Importantly, this PH symmetry holds only when both conditions $\epsilon_d =-U/2$ and $n_g = N_0+1/2$ hold. It is easy to check that also $Q_z$ in Eq.~\eqref{hamiltonianlehur} is odd under PH symmetry. Table \ref{Table:sw} summarizes the PH transformation of the various operators in Eq.~1 as well as other perturbations that we discuss next.

\subsection{Schrieffer-Wolff (SW) transformation}
\label{section:sw}
We expand the Hamiltonian at low temperatures $T \ll E_c,U$ around the two possible box charge states $N_0, N_0+1$, and around the two spin states of the small dot, using the SW transformation~\cite{hewson_1993}. To second order one generates the terms that appear in Table \ref{Table:sw}, which consist of the Hamiltonian
\bea
\label{SWpert}
H&=&H_0+\delta H(J,V_z,V_\perp,Q_z,Q_\perp)+\delta H_{ps}(\phi)  \\
&+&\Delta H_{B_f}(B_f)+\Delta H_{B}(B)+\delta H_-(J_-,V^z_-,Q^z_-,\phi_-).\nonumber
\eea
Here, $\delta H$ was introduced in Eq.~\eqref{hamiltonianlehur}. $\Delta H_{B_f} = B_f T^z$ describes a flavor field, namely an energy difference between the charge states of the box, \be
\label{eq:Bf}
B_{f}=E_c[(1-n_g)^2-n_g^2]=E_c(1-2n_{g}),
\ee
(to zeroth order in the tunneling). We also included a magnetic field term $\Delta H_{B}=B S^z,$ which requires spin-symmetry breaking. $\delta H_{ps}(\phi)=\phi \psi^\dagger \psi$ is a potential scattering term. 

All terms in $\delta H_-$ break channel symmetry. We refer to the channel-symmetry of exchanging the lead and box as a left-right (LR) symmetry, see Table.~\ref{Table:sw}. This symmetry also takes $T^z  \to -T^z$, and is satisfied in Eq.~(1). Among the terms that break this symmetry in  $\delta H_-$, specifically $J_{-}$ describes an asymmetry in the spin-Kondo interaction.

\begin{table*}[]
\begin{tabular}{|c|c|c|c|c|c|}
\hline
Interaction $\mathcal{O}_j$ & Coupling & LR symmetry & PH symmetry & Role & EK form \\ \hline
$\psi^{\dagger}\frac{\vec{\sigma}}{2}\psi\vec{S}$ &
$J$ &  + & + & $SO(5)$ & $\approx id_{-}\chi_{+}$ \\ \hline
$T^{z}\psi^{\dagger}\frac{\tau^{z}}{2}\psi$ &
$V_{z}$ &  + &+ &  $SO(5)$ &    \\ \hline
$\psi^{\dagger}\frac{\vec{\sigma}}{2}(\tau^{-}T^{+}-\tau^{+}T^{-})\psi\vec{S}$ & $Q_{\perp}$ & + & + & $SO(5)$             & $\approx id_{+}a_{-}$         \\ \hhline{|=|=|=|=|=|=|}
$\psi^{\dagger}\frac{\tau^{+}T^{-}-\tau^{-}T^{+}}{2}\psi$                        & $V_{\perp}$ & + & - & $\rightarrow SU(4)$ & $\approx i a_{+}\chi_{-}$      \\ \hline
$T^{z}\psi^{\dagger}\vec{\sigma}\tau^{z} \psi \vec{S}$               & $Q_{z}$     & + & - & $\rightarrow SU(4)$ & $\approx ia_{+}\chi_{-}$ \\ \hhline{|=|=|=|=|=|=|}
$\psi^{\dagger}\psi$ &  $\phi$ &  + & - &
\begin{tabular}[c]{@{}c@{}}marginally \\ irrelevant\end{tabular} & 
\\ \hhline{|=|=|=|=|=|=|}
$\psi^{\dagger}\frac{\vec{\sigma}\tau^{z}}{2}\psi\vec{S}$ & $J_{-}$ & - & + & marginal &
$\approx id_{+}\chi_{-}$ \\ \hline
$T^{z}\psi^{\dagger}\vec{\sigma}\psi \vec{S}$ & $Q_{z}^{-}$ & - & - & irrelevant & \\ \hline
$\frac{1}{2}T^{z}\psi^{\dagger}\psi$ &  $V_{z}^{-}$ &
- &  + & irrelevant & 
\\ \hline
$\psi^{\dagger}\tau^{z}\psi$ & $\phi_{-}$ &
- & - & marginal & $\approx ia_{-}a_{+}$ \\ \hhline{|=|=|=|=|=|=|}
$T^{z}$ & $B_{f}$ & - &  - & marginal &
$\approx ia_{-}a_{+}$ \\ \hline
\end{tabular}
\caption{Summary of interaction terms, their couplings, their LR and PH symmetries, their RG role with respect to the SO(5) fixed point, and their Majorana fermion form.}
\label{Table:sw}
\end{table*}

We relegate full explicit expressions of the coupling constants to Appendix \ref{App:SW}. PH-even couplings such as $J,Q_\perp$ and $V_z$ are generically finite, and specifically at the PH symmetric point $\epsilon_d =-U/2, n_g = 1/2$, are given by
\bea
J&=&\frac{4}{U}\left(t_{L}^{2}+ \frac{t_{B}^{2}(E_{c}+U/2)}{2E_{c}+U/2}\right),
~~~Q_\perp =4\frac{t_{L}t_{B}}{U},\nonumber\\
V_z&=&4\frac{E_{c}t_{B}^{2}}{U(2E_{c}+U/2)}.\nonumber
\eea
On the other hand the PH-odd couplings  vanish at the PH symmetric point, for example $V_\perp$ takes the form
\be
V_\perp=-t_L t_B \frac{E_c(2n_g-1)+2\epsilon_d+U}{(\epsilon_d+E_c(2n_g-1))(\epsilon_d+U)}.
\ee

\subsection{RG equations and emergent symmetries}
\label{rgeq1}

In this subsection we consider the Hamiltonian $H+\delta H(J,V,Q)$. The analysis of the remaining terms such as channel asymmetry, magnetic and flavor fields, is postponed to Sec. V. 

\begin{figure}
\includegraphics[width=6cm]{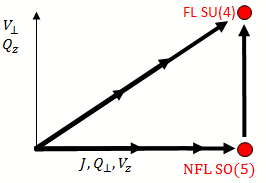}
\caption{Schematic RG flow for the PH-even couplings (horizontal axis) and PH-odd couplings (vertical axis). In the infrared, the couplings flow to one of two fixed points: a NFL phase or a FL phase. A crossover from the former to the latter occurs as the PH-odd couplings are turned on}
\label{phase2}
\end{figure}

We have seen that away from the PH symmetric point $(n_{g},\eta)=(\frac{1}{2},0)$, generically all coupling constants $J,V_z,V_\perp,Q_z,Q_\perp$ in Eq.~\eqref{hamiltonianlehur} are finite. They satisfy the following weak-coupling RG equations~\cite{le2004maximized}:
\bea
\frac{dJ}{d l} &=&  J^2+Q_z^2+2 Q_\perp^2,\nonumber \\
\frac{dV_z}{d l} &=&  V_\perp^2+3 Q_\perp^2,\nonumber \\
\frac{dV_\perp}{d l} &=& V_\perp V_z + 3 Q_\perp Q_z  ,\nonumber \\
\frac{dQ_z}{d l} &=& 2 J Q_z+2V_\perp Q_\perp  ,\nonumber \\
\frac{dQ_\perp}{d l} &=&  2 J Q_\perp+V_z Q_\perp+V_\perp Q_z.
\label{LehureRG}
\eea
Here $l$ is the logarithmic RG scale parameter.
In a similar fashion to the Kondo RG equations displaying the irrelevancy of spin-anisotropy, it was noticed ~\cite{borda20034,le2004maximized} that the present system of RG equations flows to a strong coupling fixed point with equal couplings. When the system reaches isotropy $V=Q=J$, these RG equations become $dJ/dl=4J^2$. The system flows to the upper fixed point in Fig.~\ref{phase2} on the Kondo scale~\cite{le2004maximized}
\be
T_K^{SU(4)} \sim D e^{-1/4J},
\ee
where $D$ is the conduction electron's band width.

Now consider the $(n_{g},\eta)=(\frac{1}{2},0)$ point where $V_\perp=Q_z=0$. A similar analysis of the isotropy of the remaining couplings $J,V_z,Q_\perp$, suggests a flow to an isotropic fixed point (see lower fixed point at Fig.~\ref{phase2}) with $dJ/dl=3J^2$, and the partially isotropic Hamiltonian flows to the lower fixed point in Fig.~\ref{phase2}, with a slightly smaller Kondo scale
\be
T_K^{SO(5)} \sim D e^{-1/3J}.
\ee

\subsubsection{SU(4) versus SO(5)}
\label{sec:su4vs}
Consider the matrix of 15 generators
\bea
&&T^{ab}=\\
&&\frac{1}{2}\begin{pmatrix}
	0 & \sigma^{3} & -\sigma^{2} & \sigma^{1}\tau^{1} & \sigma^{1}\tau^{2} & \sigma^{1}\tau^{3} \\
	-\sigma^{3} & 0 & \sigma^{1} & \sigma^{2}\tau^{1} & \sigma^{2}\tau^{2} & \sigma^{2}\tau^{3} \\
	\sigma^{2} & -\sigma^{1}  & 0 & \sigma^{3}\tau^{1}  &\sigma^{3}\tau^{2} & \sigma^{3}\tau^{3}\\
	-\sigma^{1}\tau^{1} & -\sigma^{2}\tau^{1}& -\sigma^{3}\tau^{1}  & 0 &\tau^{3} & -\tau^{2}\\
	-\sigma^{1}\tau^{2} & -\sigma^{2}\tau^{2} & -\sigma^{3}\tau^{2} & -\tau^{3} & 0 & \tau^{1}\\
	-\sigma^{1}\tau^{3} &- \sigma^{2}\tau^{3} &- \sigma^{3}\tau^{3}  & \tau^{2}& -\tau^{1} & 0 \nonumber
\end{pmatrix}.
\label{Tmatrix}
\eea
The 15 independent generators appearing in the upper diagonal part  ($T^{ab}=-T^{ba}$ with $a,b=1...6$) satisfy  the algebra $[T^{ab},T^{cd}]=-i (\delta_{bc} T^{ad}-\delta_{ac} T^{bd}-\delta_{bd} T^{ac}+\delta_{ad} T^{bc})$. Hence~\cite{georgi2018lie} they are 15 generators of SO(6), which is isomorphic to SU(4). 
Specifically, these matrices form a 4 dimensional representation.

We can  organize the impurity operators $S$ and $T$ into generators of this SO(6) symmetry. One introduces a singly occupied fermionic site carrying spin and flavor indices~\cite{itoi2000phase}, $f^{\dagger}_{\alpha \sigma} f_{\alpha \sigma}=1$, in terms of which $S^a=f^\dagger \frac{\sigma^a}{2} f$, $T^b=f^\dagger \frac{\tau^b}{2} f$ and $2 S^a T^b=f^\dagger \frac{\sigma^a \tau^b}{2} f$. Then the 4 impurity states form a representation of SO(6)
\bea
&&f^\dagger T^{ab} f=M^{ab}=\\
&&\begin{pmatrix}
	0 & S^z & -S^y & 2 S^x T^x& 2S^x T^y & 2S^x T^z \\
	-S^z & 0 & S^x & 2S^yT^x & 2S^yT^y & 2S^yT^z \\
	S^y & -S^x  & 0 & 2S^z T^x  &2S^z T^y & 2S^zT^z\\
	-2S^xT^x & -2S^yT^x& -2S^zT^x  & 0 &T^z & -T^y\\
	-2S^xT^y & -2S^yT^y & -2S^zT^y & -T^z & 0 & T^x\\
	-2S^xT^z &- 2S^yT^z &- 2S^zT^z  & T^z& -T^x & 0 \nonumber
\end{pmatrix}.
\label{Tmatriximp}
\eea
Evidently, the fully isotropic situation with $J=V_z=V_\perp=Q_z=Q_\perp$ allows us to write Hamiltonian $H+\delta H$ in Eq.~1 in SO(6) [or equivalently SU(4)] isotropic form
\begin{align}
\label{eq:Hso6}
H_{\rm SO(6)} = H_0+ J \sum_{A=1}^{15} J^A M^A \;,
\end{align}
where $J^A=\psi^{\dagger} T^A \psi$  and $\sum_{A=1}^{15}=\sum_{a<b,a,b=1}^6$.

The case with PH symmetry in our model corresponds to $V_\perp= Q_z=0$. Then the Hamiltonian Eq.~1 can be written in terms of the 10 generators of SO(5), which is a subgroup of SO(6). These 10 generators are given by $T^{ab}$ with $a,b=1...5$, (given by the matrix in Eqs.~(\ref{Tmatrix}) by removing the last column and row), 
\begin{align}
\label{eq:Hso5}
H_{\rm SO(5)} = H_0+ J \sum_{A=1}^{10} J^A M^A \;,
\end{align}
where $\sum_{A=1}^{10}=\sum_{a<b,a,b=1}^5$.

\subsubsection{Generalized anisotropic RG equations}
The RG equations Eq.~\eqref{LehureRG} assumed spin SU(2) symmetry as reflected e.g. in the  Kondo coupling $J$. Similarly it assumed flavor U(1) symmetry implying $V_{x}=V_{y}$ and $Q_{x}=Q_{y}$. Motivated by the next chapter based on the anisotropic Emery-Kivelson approach, in which the spin SU(2) symmetry is broken, we now generalize the RG equations to the fully anisotropic case of the form $H_0+\sum_{a<b, a,b=1}^6\lambda_{ab}   (\psi^{\dagger}T^{ab}\psi)(f^{\dagger}T^{ab}f)$. We associate each of these terms with a coupling $\lambda_{ab}$
\begin{equation}
\lambda_{ab} =
\begin{pmatrix}
0 & J_{z} & J_{y} & Q_{x}^{x} & Q_{y}^{x} & Q_{z}^{x}\\
J_{z} & 0 & J_{x} & Q_{x}^{y} & Q_{y}^{y} & Q_{z}^{y}\\
J_{y} & J_{x} & 0 & Q_{x}^{z} & Q_{y}^{z} & Q_{z}^{z}\\
Q_{x}^{x} & Q_{x}^{y} & Q_{x}^{z} & 0 & V_{z} & V_{y}\\
Q_{y}^{x} & Q_{y}^{y} & Q_{y}^{z} & V_{z} & 0 & V_{x}\\
Q_{z}^{x} & Q_{z}^{y} & Q_{z}^{z} & V_{y}&  V_{x} & 0
\end{pmatrix}
.
\label{eq:lambda}
\end{equation}
Here, we defined $Q_b^a$ by generalizing the spin-flavor coupling terms $Q_b$ in Eq.~(1) to 
$Q_b^a \psi^\dagger \sigma^a \tau^b \psi S^a T^b$, yielding 15 independent coupling constants.
As demonstrated in the Appendix, these 15 coupling constants satisfy the RG equation
\bea
\label{eq:ope}
\frac{d\lambda_{ij}}{dl}=\sum_{k \ne i,j}\lambda_{ik}\lambda_{kj}.
\eea
For example, the RG flow of $J_{x}$  is given by
\bea
\label{rgequations}
\frac{dJ_{x}}{dl}=\frac{d\lambda_{23}}{dl}&=&\lambda_{21}\lambda_{13}+\lambda_{24}\lambda_{43}+\lambda_{25}\lambda_{53}+\lambda_{26}\lambda_{63} \nonumber \\ 
&=&J_{z}J_{y}+Q_{x}^{y}Q_{x}^{z}+Q_{y}^{y}Q_{y}^{z}+Q_{z}^{y}Q_{z}^{z}. 
\eea
This system of equations is numerically solved in appendix \ref{se:numericalRG}, showing that the isotropic fixed point is achieved for anisotropic bare values including spin anisotropy, which corresponds to the Toulouse limit discussed in the next section.

\section{Mapping to a model of Majorana fermions}

\subsection{Bosonization and refermionization}
\label{Sec:EK}
In this section we apply the techniques that Emery and Kivelson (EK) employed to the spin-2CK model~\cite{emery1992mapping}, to solve our spin-flavor model Eq.~\eqref{hamiltonianlehur}.  The resulting model consists of a fixed point Hamiltonian which is quadratic in terms of bulk and local Majorana fermions. While in this chapter we present the mapping to the various terms of the Hamiltonian into the Majorana description, in Sec.~IV we will apply this approach to construct the phase diagram.

As discussed in detail in Refs.~\onlinecite{zarand1998simple,von1998bosonization}, the starting point is to write the free part of the Hamiltonian in terms of chiral one-dimensional fermions, $H_{0}=\sum_{\alpha,\sigma}iv_{F}\int^{\infty}_{-\infty}\frac{dx}{2\pi}\psi_{\alpha \sigma}^{\dagger}\partial_{x}\psi_{\alpha \sigma}$ where $v_F$ is the Fermi velocity. We now treat the various perturbations. We first consider the terms $H_J+H_{V_z}$, i.e.,  (i) the spin-Kondo interactions in Eq.~(1) which we assume to be anisotropic
\bea
H_{J}&=&\frac{1}{2}J_{z}S^{z}(\psi_{ \alpha\uparrow}^{\dagger}\psi_{ \alpha\uparrow}-\psi_{\alpha\downarrow}^{\dagger}\psi_{\alpha\downarrow})\\
&+&\frac{1}{2}J_{\perp}(S^{+}\psi_{ \alpha\downarrow}^{\dagger}\psi_{ \alpha\uparrow}+S^{-}\psi_{ \alpha\uparrow}^{\dagger}\psi_{ \alpha\downarrow}),\nonumber
\eea
together with (ii) the $V_z$ term $H_{V_z}=\frac{1}{2}V_{z}T^{z}\psi^\dagger \tau^z \psi$.

The EK transformation begins with the replacement of the fermionic fields 
$\psi_{\alpha\sigma}(x)$ with  bosonic fields $\phi_{\alpha\sigma}(x)$, using the relation,
\bea
\label{eq:psiphi}
\psi_{\alpha\sigma}(x)=F_{\alpha\sigma}a^{-1/2}e^{-i\phi_{\alpha\sigma}(x)}.
\eea
Here $a$ is a short distance cutoff which we set to unity. The Klein factors $F_{\alpha\sigma}$  retain the fermionic 
commutation relations. The spin-flip Kondo term becomes
\bea
H_{j_{\perp}}=\frac{1}{2}J_{\perp}(S^{+}F^{\dagger}_{\alpha\downarrow}F_{\alpha\uparrow}e^{i\phi_{\alpha\downarrow }}e^{-i\phi_{ \alpha\uparrow}}+h.c.).
\eea
We  then use the bosonization  identity\cite{von1998bosonization} $
\psi^{\dagger}_{\alpha \sigma}(x)\psi_{\alpha \sigma}(x)=\partial_{x}\phi_{\alpha\sigma}(x)$ to bosonize the $J_{z}$ term,
\bea
H_{J_{z}}&=&\frac{1}{2}J_{z}S^{z}(\partial_{x}\phi_{ \alpha\uparrow}-\partial_{x}\phi_{\alpha\downarrow}).
\eea
Likewise the free Hamiltonian maps to $H_{0}=\sum_{\alpha \sigma}
v_{F}\int^{\infty}_{-\infty}\frac{dx}{2\pi}\frac{1}{2}(\partial_{x}\phi_{\alpha \sigma})^{2}$. Next, the bosonic fields $\phi_{\alpha \sigma}$ are re-expressed in a basis of charge, spin, flavor and spin-flavor degrees of freedom, denoted by $\phi_{\mathcal{A}}$ ($\mathcal{A}=c,s,f,sf$),
\bea
\label{eq:csfsf}
\phi_{c}=\frac{1}{2}(\phi_{1\uparrow}+\phi_{1\downarrow}+\phi_{2\uparrow}+\phi_{2\downarrow}),\\
\phi_{s}=\frac{1}{2}(\phi_{1\uparrow}-\phi_{1\downarrow}+\phi_{2\uparrow}-\phi_{2\downarrow}),\nonumber\\
\phi_{f}=\frac{1}{2}(\phi_{1\uparrow}+\phi_{1\downarrow}-\phi_{2\uparrow}-\phi_{2\downarrow}),\nonumber\\
\phi_{sf}=\frac{1}{2}(\phi_{1\uparrow}-\phi_{1\downarrow}-\phi_{2\uparrow}+\phi_{2\downarrow}).\nonumber
\eea
Thus
\bea
H_{0}&+&H_{J}=v_{F}\int^{\infty}_{-\infty}\frac{dx}{2\pi}\frac{1}{2}(\partial_{x}\phi_{s})^{2}
+J_{z}S^{z}\partial_{x}\phi_{s}(0)\\
&+&\frac{1}{2}J_{\perp}S^{+}e^{-i\phi_{s}}[F^{\dagger}_{1\downarrow }F_{1 \uparrow}e^{-i\phi_{sf}(0)}+F^{\dagger}_{2 \downarrow }F_{2 \uparrow}e^{+i\phi_{sf}(0)}]+h.c\nonumber
\eea
We proceed to perform unitary rotations generated by the $S^{z}$ and $T^{z}$ operators,
\be
U=e^{i(\gamma_{s}S^{z}\phi_{s}+\gamma_{f}T^{z}\phi_{f})}.
\ee
In the spin-flip term $J_\perp$, the spin ladder operators  transform as
$S_{\pm}\rightarrow S_{\pm}e^{\pm i \gamma_s \phi_{s}}$, yielding
\bea
\label{ekrotate}
H_{0}&+&H_{J}+H_{V_z}\rightarrow \nonumber \\
H_{0}&+&
\frac{1}{2}J_{\perp}(S^{+}e^{i\gamma_{s}\phi_{s}})e^{-i\phi_{s}}[F^{\dagger}_{1\downarrow}F_{ 1\uparrow}e^{-i\phi_{sf}}\nonumber\\
&+&F^{\dagger}_{2\downarrow}F_{ 2\uparrow}e^{+i\phi_{sf}}]+h.c\\
&+&(J_{z}-\gamma_{s} v_{F})S^{z}\partial_{x}\phi_{s}\nonumber\\
&+&(V_{z}-\gamma_{f} v_{F})T^{z} \partial_{x}\phi_{f}.\nonumber
\eea
A particular choice of the anisotropy  parameters $J_z$ and $V_z$, satisfying
\be
\label{Toulousepnt}
\gamma_{s}=\frac{J_{z}}{v_{F}},~~~ \gamma_{f}=\frac{V_{z}}{v_{F}},~~~({\rm{Toulouse~point}}),
\ee
results in a cancellation of the two last lines in Eq.~(\ref{ekrotate}). In addition, the vertex parts $e^{-i \phi_\mathcal{A}}$ of both spin and flavor sectors cancel in all Kondo interactions, including those of the $Q$ and $V$ operators, if $\gamma_{s}=\gamma_{f}=1$.

The spin-flip term becomes
\bea
\frac{1}{2}J_{\perp}S^{+}[F^{\dagger}_{ 1\downarrow}F_{ 1\uparrow}e^{-i\phi_{sf}}+F^{\dagger}_{ 2\downarrow}F_{2\uparrow}e^{+i\phi_{sf}}]+h.c..
\eea
We proceed to define new Klein operators expressed in the basis of $\mathcal{A}=c,s,f,sf$, satisfying the following relations~\cite{zarand1998simple}
\bea
\label{zarand1998simple11}
F_{x}^{\dagger}F_{s}^{\dagger}\equiv F_{\uparrow 1}^{\dagger}F_{\downarrow 1}, \ \ F_{x}F_{s}^{\dagger}\equiv F_{\uparrow 2}^{\dagger}F_{\downarrow 2},\\
F_{x}^{\dagger}F_{f}^{\dagger}\equiv F_{\uparrow 1}^{\dagger}F_{\uparrow 2}, \ \ F_{c}^{\dagger}F_{s}^{\dagger}\equiv F_{\uparrow 1}^{\dagger}F_{\uparrow 2}^{\dagger}.\nonumber
\eea
Then the spin-flip term becomes
\bea
H_{J_\perp}=\frac{1}{2}J_{\perp}S^{+}[F_{s}(F_{sf}e^{-i\phi_{sf}}+F^{\dagger}_{sf}e^{i\phi_{sf}})]+h.c.
\eea
Finally we refermionize, i.e. rewrite these interactions in terms of the new fermionic operators (either local- or bulk fermions) 
\bea
\psi_{\mathcal{A}}(x)&=& F_{\mathcal{A}}e^{-i\phi_{\mathcal{A}}(x)}, ~~~(\mathcal{A}=c,s,f,sf),\\  d&=&F_{s}^{\dagger}S_{-},~~~a=F^{\dagger}_{f}T_{-},
\eea
where $\psi_{\mathcal{A}}$ are bulk fermions and $d,a$ are local fermions, corresponding to the spin and flavor impurity degrees of freedom, respectively. The bulk fermions can be used to define Majorana fermion fields evaluated at $x=0$,
\bea
\label{eq:majorana}
\chi_{\mathcal{A}+}\equiv\frac{\psi_{\mathcal{A}}^{\dagger}+\psi_{\mathcal{A}}}{\sqrt{2}},~~~\chi_{\mathcal{A}-}\equiv\frac{\psi_{\mathcal{A}}^{\dagger}-\psi_{\mathcal{A}}}{i\sqrt{2}}.
\eea
We also define  local Majorana fermions (``Majoranas") 
\bea
\label{eq:majoranalocal}
a_{+}\equiv\frac{a^{\dagger}+a}{\sqrt{2}},~~~a_{-}\equiv\frac{a^{\dagger}-a}{i\sqrt{2}},
\eea
and similarly for $d_\pm$. Notice that $a^\dagger a=\frac{1}{2}+T^z=\frac{1}{2}+ia_- a_+$ and $a_\pm^2=\frac{1}{2}$ (and similarly for $d_\pm$). 

Thus, after the EK transformation we  ended up with (i) 4 local Majoranas, accounting for the $2 \log 2$ impurity entropy of the free fermion fixed point, due to  the spin and flavor impurity degrees of freedom $S$ and $T$; (ii) 8 Majorana fields $\chi_{\mathcal{A} \pm}$. Our $J_{\perp}$ term, for example, takes 
the form $i J_{\perp}d_{-}\chi_{sf+}(0)$, which couples the local spin Majorana ($d_{-}$) and the conduction electrons' spin-flavor degree of freedom $\chi_{sf+}$.

\subsection{Toulouse limit of the SO(5) fixed point}
The PH symmetric model which flows to the SO(5) symmetric point is obtained by combining the terms $\mathcal{H}^*_{SO(5)}=H_0+H_J+H_{V_z}+H_{Q_\perp}$. At the Toulouse point, it is given by
\bea
\label{Ekhamiltonian1}
\mathcal{H}^*_{SO(5)}&=&H_{0}+iJ_{\perp}d_{-}\chi_{+}-2Q_{\perp}^{z}a_{-}\chi_{+}d_{-}d_{+}\\
&-&2iQ_{\perp}^{\perp}d_{+}a_{-},\nonumber 
\eea
where $\chi_{\pm} \equiv \chi_{sf\pm}(x=0)$.

We now consider the Hamiltonian Eq.~\eqref{Ekhamiltonian1} from a RG perspective. With respect to the free-fermion fixed point $H_0$,  bulk fermions such as $\chi_\pm$ have scaling dimension $\Delta_\chi=1/2$ (with correlation function decaying as $1/t^{2 \Delta_\chi}$), while  local Majoranas have scaling dimension $0$. Boundary operators of scaling dimension $\le 1$ are relevant. 

Firstly, the $J_{\perp}$ term resulting from the spin-flip interaction is relevant ($\Delta_{J_{\perp}} =1/2$). At energies lower than the Kondo temperature $\sim J_\perp^2$ the Majorana $d_-$ gets ``absorbed" into the Majorana bulk field $\chi_+$, resulting in~\cite{sela2009nonequilibrium}
\be
\label{replacement}
d_{-}\approx\frac{\chi_{+}}{\sqrt{T_{K}}}.
\ee
Thus $d_-$ picks up a scaling dimension $1/2$ near the Kondo fixed point.

The  $Q_{\perp}^{\perp}$   operator of dimension 0  gaps out the $d_{+}$ and $a_{-}$ operators. The product $i a_{-}d_{+}$ obtains then an expectation value. Replacing this operator in  $Q_{\perp}^{z}$ by a constant  results in a term of the same form as $\chi_{+}d_{-}$, which can be used to redefine $J_{\perp} \to J_{\perp}'$. We end up with a Majorana  fixed point Hamiltonian, whose couplings are represented visually in the figure in column 1 of Table.~\ref{Tb:1},
\bea
\label{Ekhamiltonian1new}
\mathcal{H}^*_{SO(5)}&\to&H_{0}+iJ_{\perp}'d_{-}\chi_{+}-2iQ_{\perp}^{\perp}d_{+}a_{-}.
\eea
\subsubsection{Perturbations}
Next, we add symmetry breaking perturbation to the Toulouse fixed point Hamiltonian.  Channel symmetry breaking represented by $J_{-}$ term in Eq.~(\ref{SWpert}) takes the form
\bea
\label{Ekhamiltonian2}
H_{J_-}&=&iJ_{-}d_{-}d_{+}\psi_{sf}^{\dagger}\psi_{sf}
-iJ_{-}d_{+}\chi_{-}.
\eea
As will be discussed in Sec.~\ref{se:NFLFP}, both terms are irrelevant since $d_+$ is already gapped out. The least irrelevant,  second term, couples $d_{+}$ to $\chi_{-}$, in addition to the couplings of Eq.~\eqref{Ekhamiltonian1new} (see figure in column 3 of Table.~\ref{Tb:1}). 

Additional key perturbations emerge from breaking PH symmetry. First, those include $V_\perp$ and $Q_z$,
\bea
\label{Ekhamiltonian3}
H_{V_\perp,Q_z}=-iV_{\perp}a_{+}\chi_{-}+2Q_{z}(a_{-}a_{+})d_{+}\chi_{-} \nonumber \\
-Q_{z}(a_{-}a_{+})d_{-}d_{+}\psi_{sf}^{\dagger}\psi_{sf}.
\eea
As before, we can use the finite expectation value of $id_+a_-$ to see that the second term $(Q_z)$ simpy renormalizes the first term $(V_\perp)$. Similarly the third term in the second line is less relevant than the first two.

Second, one has the flavor field perturbation
\be
H_{B_f}=iB_f' a_- a_+.
\ee
In addition to the bare flavor field $E_c(1-2n_g)$, this operator corrects the coefficient of $B_f$ due to higher order terms including $\phi_-$. We thus consider this as a renormalization of $n_g$.

\subsubsection{Deviations from the Toulouse point}

Generically there are deviations from the Toulouse point condition Eq.~(\ref{Toulousepnt}), resulting in terms of the form
\bea
\label{Toulouse}
\delta H_{s}&=&i(J_{z}- v_{F})d_{-}d_{+}\psi_{s}^{\dagger}\psi_{s},\\
\delta H_{f}&=&i(V_{z}-v_{F})a_{-}a_{+}\psi_{f}^{\dagger}\psi_{f}.\nonumber
\eea
As we shall see in the following section, these  terms are important for the analysis of the effect of generic perturbations on transport and thermodynamic properties of the system.

\section{NFL SO(5) fixed point}
\label{se:NFLFP}
The PH symmetric NFL SO(5) point is located at $n_{g}=1/2$ and  $\eta=0$, where in addition a channel symmetry is imposed  by tuning the tunnelings $t_{L},t_{B}$ so that $J_-=0$. Then the effective Hamiltonian is given by Eq.~\eqref{Ekhamiltonian1new}. As summarized in column 1 of Table.~\ref{Tb:1}, in this section we present the thermodynamic behavior in terms of entropy, spin and flavor susceptibilities, and the conductance at the SO(5) point.

\newcommand{\addpica}{\includegraphics[width=6em]{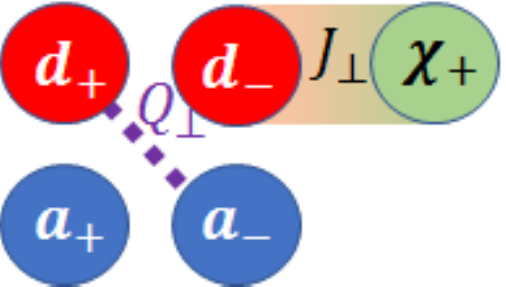}}
\newcommand{\addpicb}{\includegraphics[width=6em]{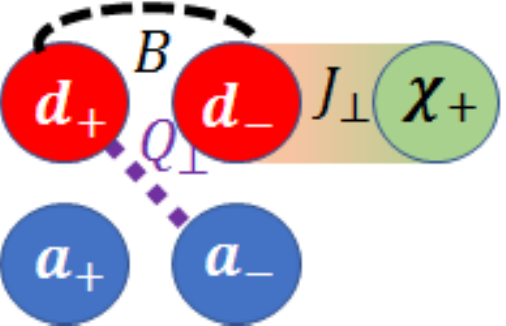}}
\newcommand{\addpicc}{\includegraphics[width=8em]{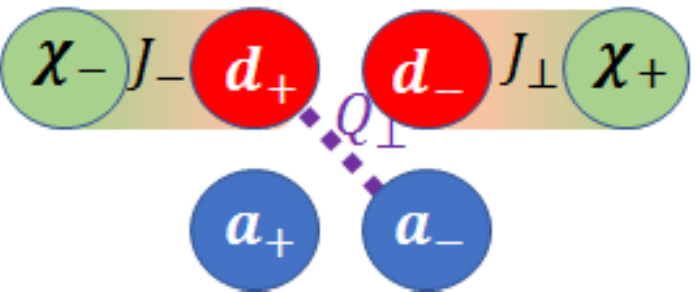}}
\newcommand{\addpice}{\includegraphics[width=6em]{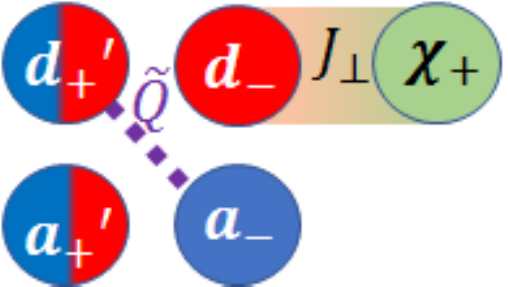}}
\newcommand{\addpicd}{\includegraphics[width=8em]{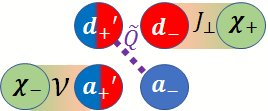}}
\newcolumntype{C}[1]{>{\centering\arraybackslash}m{#1}}
\begin{table*}
	\renewcommand{\arraystretch}{1.5}
	\centering
	\begin{tabular}{|C{1.5cm}|C{2cm}|C{2cm}|C{3cm}|C{3cm}|C{2cm}|}
		\hline
		&\ \ \ \ (1): \newline SO(5) symmetry & \ \ \ \ (2): \newline magnetic field $B$ &\ \ \ \ \ (3): \newline channel asymmetry $J_-$&\ \ \ \ \ (4):\newline  PH breaking I~~ ($B_f$, $V_{\perp}$) &\ \ \ \  (5):\newline PH breaking II -NFL line \\ \hline
		$S_{imp}$ & $\frac{1}{2}\log{2}$ & $\frac{1}{2}\log{2}$ & $\frac{1}{2}\log{2}$ & 0 & $\frac{1}{2}\log{2}$ \\ \hline
		$\chi_s$ & $\omega$ & $\omega$ &  $\omega$ &  $\omega$ & $const.$\\ \hline
		$\chi_f$ & $\omega$ & $const.$ & $const.$ & $\omega$ & $const.$\\\hline 
		$\delta G (\omega)$ & $(\frac{\omega}{T_{k}})^{\frac{3}{2}}$ & $(\frac{\omega}{T_{k}})^{\frac{1}{2}}$ & $(\frac{\omega}{T_{k}})^{\frac{1}{2}}$ & $(\frac{\omega}{T_{k}})$ & $(\frac{\omega}{T_{k}})^{\frac{1}{2}}$\\\hline
		Majorana coupling scheme & \addpica & \addpicb & \addpicc & \addpicd &\addpice \\ \hline
	\end{tabular}
	\caption{A summary of characteristics of different states of our spin-flavor model (rows/columns correspond to different observables/states). The system's states are (from left  to right column): \textbf{(1) SO(5) symmetry} with both PH symmetry ($n_{g}=\frac{1}{2}$, $\eta=0$) and LR symmetry ($t_L$ and $t_B$ tuned such that $J_{-}=0$). \textbf{(2) Broken spin symmetry} with an external magnetic field $B$. \textbf{(3) Channel asymmetry}: the tunneling couplings $t_{L},t_{B}$ are detuned such that $J_{-}\neq 0$. \textbf{(4) PH breaking I:} Upon adding generic PH-asymmetry ($\epsilon_d/U \ne -1/2$, $n_g \ne 1/2$), the system becomes a FL. \textbf{(5) PH breaking II:} PH symmetry breaking parameters are fine-tuned such that a NFL is restored.
		The rows, from top to bottom, stand for:  $T \to 0$ impurity's entropy,  spin susceptibility,  flavor susceptibility and  conductance between source and drain leads. The final row illustrating the couplings between the local and bulk Majorana fermions. The red/blue/green balls represent local-spin/local-flavor/bulk Majorana fermions. Dashed lines connect local Majoranas, while reddish strips represent couplings of a local Majoranas to a Majorana field. Mixed color balls imply that a Majorana basis rotation was performed.}
	\label{Tb:1}
\end{table*}

\subsection{Entropy}
Looking at $\mathcal{H}^*_{SO(5)}$ in Eq.~\eqref{Ekhamiltonian1new}, the Hamiltonian with both channel and PH symmetries, we see that out of the 4 local Majoranas, only $a_{+}$ is decoupled (see schematic illustration in column 1 in Table.~\ref{Tb:1}).

Similar to the spin-2CK case with an unpaired Majorana~\cite{emery1992mapping}, this explains the fractional entropy of $\frac{1}{2}\log{2}$, observed for $T\rightarrow 0$ in either one of the NRG plots in the first row in Fig.~\ref{graph_table1} where the perturbations are sent to zero [LR symmetry $\lambda=1$ in Fig.~\ref{graph_table1}(a), zero magnetic field $B=0$ in Fig.~\ref{graph_table1}(d), or PH symmetry $\eta=\epsilon_d+U/2=0$ in Fig.~\ref{graph_table1}(g)]. In contrast to the spin-2CK model, in our case the free Majorana is assigned to the flavor degree of freedom $T$.

\subsection{Spin and flavor susceptibilities}

\label{so5sus}
Consider the spin susceptibility $\chi_{s}(t)=\langle S^{z}(t)S^{z}(0)\rangle$. The impurity's spin can be written in the Majorana language as $S^{z}=d^{\dagger}d-\frac{1}{2}=id_{-}d_{+}$ using Eq.~\eqref{eq:majoranalocal}. From Eq.~(\ref{replacement}) we then obtain
\bea
\label{spinsus}
\chi_{s}(t) \propto \frac{1}{T_{K}} \langle \chi_{+}(t)\chi_{+}(0)\rangle\langle d_{+}(t)d_{+}(0)\rangle .
\eea
For the dimension-$1/2$ fermion field $\chi_{+}$, we have $\langle\chi_{+}(t)\chi_{+}(0)\rangle\propto\frac{1}{t}$. We now study the behavior of $\langle d_{+}(t)d_{+}(0) \rangle$.

For comparison, in the spin-2CK model the Majorana $d_+$ is decoupled and thus has no dynamics, in which case $\langle d_{+}(t)d_{+}(0)\rangle=1/4$ and $\chi_{s}^{spin-2CK}(t)\propto\frac{1}{t}$; Fourier transforming, $\chi_{s}^{spin-2CK} (\omega) \propto {\rm{const}}$. In other words, the impurity spin operator admits a low energy expansion in terms of a dimension-$1/2$ field
\be
\label{eq:spin-2CK}
S^z_{spin-2CK} \sim \frac{i}{\sqrt{T_K}} \chi_+ d_+ ~~~~~({\rm{scaling~dimension~1/2}}). 
\ee
In our spin-flavor Kondo model, to find $\langle d_{+}(t)d_{+}(0)\rangle$ we recall that  $d_{+}$ is gapped out along with $a_{-}$ by the $Q_{\perp}^\perp$ term in Eq.~\eqref{Ekhamiltonian1new}. This relevant term separates the Hilbert space into low energy and high energy sub-spaces, with energy difference $2Q_\perp^\perp$, which are the $\pm 1/2$ eigenspaces of $i  d_+ a_-$. Equivalently we define a complex fermion $f=\frac{1}{\sqrt{2}}(a_{-}+id_{+})$, so that $f^\dagger f=\frac{1}{2}+i a_-d_+$ and $d_+=\frac{f-f^\dagger}{\sqrt{2}i }$. Namely, $H_{Q_\perp}=-2i Q_\perp^\perp d_+ a_-=2 Q_\perp^\perp (f^\dagger f-\frac{1}{2})$. The ground state subspace $i a_- d_+  =-1/2$ corresponds to $f^\dagger f=0$, and the excited subspace $i a_-d_+ =1/2$ to $f^\dagger f=1$. We define projectors into these subspaces
\bea
\label{projectors}
\mathcal{P}_{0}=1-f^{\dagger}f,~~~ \mathcal{P}_{1}=f^{\dagger}f.
\eea
The operators $f^{\dagger}$ ($f$) appearing in $S^z$ bridge the low and high energy subspaces. Therefore, $\mathcal{P}_{0}S^{z} \mathcal{P}_{0}=\mathcal{P}_{1}S^{z} \mathcal{P}_{1}=0$, meaning that $\langle d_{+}(t)d_{+}(0)\rangle$ strongly oscillates as $e^{i  2Q_\perp^\perp t}$. Similar to a SW transformation, to obtain the leading operator expansion of $S^z$ at the spin-flavor fixed point, we  perform perturbation theory in off-diagonal operators with respect to projectors $\mathcal{P}_{01}$. We define for any operator $\mathcal{O}_{ij}=\mathcal{P}_i \mathcal{O} \mathcal{P}_j$, including for the Hamiltonian $H_{ij}=\mathcal{P}_i H \mathcal{P}_j$. For the fixed point Hamiltonian Eq.~(\ref{Ekhamiltonian1new}), off-diagonal terms $H_{01}$, $H_{10}$ do exist and originate from deviations from the Toulouse point, see $\delta H_{s,f}$ in Eq.~\eqref{Toulouse}, since they include only one Majorana out of the pair $d_+, a_-$. To second order in off diagonal operators, we find that the expectation value of any off diagonal operator (like $S^z$) takes the form $\langle \mathcal{O} \rangle = \langle \mathcal{O}_{01} (E-H_{11})^{-1}  H_{10}+H_{10}(E-H_{11})^{-1} \mathcal{O}_{10} \rangle $ where $E$ here is the ground state energy. Using $E-H_{11} \cong -2Q_\perp^\perp$ for the excitation energy, we see that off diagonal operators do obtain an effective operator form acting in the ground state subspace
\be
\label{eq:pert_theo_op}
\mathcal{O}_{00,eff}=\frac{1}{-2Q_\perp^\perp} (\mathcal{O}_{01}  H_{10}+H_{10} \mathcal{O}_{10}).
\ee
Thus, we obtain
\be
\label{eq:Szsf}
S^z_{spin-flavor} \sim -\frac{J_z-v_F}{4Q^{\perp}_\perp } \psi^\dagger_s \psi_s- i \frac{V_z-v_F}{2Q^{\perp}_\perp\sqrt{T_{K}}} \chi_+ \psi^\dagger_f \psi_f a_+ + ....
\ee
The first term has a scaling dimension $1$, and coincides with the spin current~\cite{affleck1990current}. This term appears in any FL state. The second terms has scaling dimension $3/2$ and the dots stand for less relevant terms.

From the first leading term, the spin susceptibility 
decays as $\frac{1}{t^{2}}$. In frequency space this becomes $\chi_s(\omega) \sim \omega$, as displayed column 1 row 2 in Table \ref{Tb:1}, matching the linear behavior of the NRG calculated spin susceptibility in Fig.~\ref{graph_table1}. 

The flavor susceptibility $\chi_{f}(t)=\langle T^{z}(t)T^{z}(0)\rangle$ can be dealt with in a similar manner using its fermionic form $T^{z}=a^{\dagger}a-\frac{1}{2}=ia_{-}a_{+}$. The leading term is 
\be
T^z_{spin-flavor} \sim -\frac{V_z-v_F}{4Q^{\perp}_\perp } \psi^\dagger_f \psi_f  +..., 
\ee
which has scaling dimension 1, (with an additional dimension 3/2 operator originating from $J_z-v_F$), leading to a FL-like behavior $\chi_{f}(t) \sim  \frac{1}{t^{2}}$, or $\chi_{f}(\omega) \sim \omega$, matching our NRG results in Fig.~\ref{graph_table1}.\\

\subsection{Leading irrelevant operator and conductance}
The temperature dependence of the conductance for the multi-channel Kondo effect probes the leading irrelevant (boundary) operator at the non-trivial fixed point~\cite{affleck1991critical,pustilnik2004quantum}.
From scaling analysis, a boundary perturbation of dimension $\Delta$ 
yields to first order a temperature dependence of the conductance of the form $\delta G \sim T^{\Delta-1}$. 

In the spin-2CK state there exists an anomalous operator of dimension $\Delta=3/2$ originally found by CFT~\cite{affleck1991critical}, leading to the experimentally observed $\delta G \sim T^{1/2}$ scaling of the conductance~\cite{potok2007observation}. We can identify this dimension $3/2$ operator in the Majorana fermion language: In the spin-2CK case, with the $Q_\perp^\perp$ term absent and hence $d_+$ being a free Majorana, $\delta H_s$ in Eq.~(\ref{Toulouse})  becomes
\bea
(\delta H_s)^{(spin-2CK)}  \sim  \frac{i (J_z-v_F)}{\sqrt{T_K}} (\chi_+ \psi^\dagger_s \psi_s) d_+,\nonumber \\
{\rm{(scaling~dimension~3/2).~~~~~~~~ ~~~~~~~}}
\eea

As discussed in Sec.~\ref{so5sus}, in the spin-flavor model $\delta H_s$ does not commute with $Q_\perp^\perp$ and is off-diagonal in our 2-subspace decomposition Eq.~(\ref{projectors}). The leading irrelevant operator can be obtained from second order perturbation theory in off-diagonal operators,
\be
\label{eq:pert_theo_Ham}
H_{00,eff}=H_{01}(E-H_{11})^{-1} H_{10}.
\ee
To obtain a nontrivial operator, this time, we have to combine $\delta H_s$ and $\delta H_f$ to obtain
\bea
\label{eq:leading53}
\delta H_{irr}=-i\frac{(J_z-v_F)(V_z-v_F)}{2Q_\perp^\perp \sqrt{T_K}} (\chi_+ \psi^\dagger_s \psi_s \psi^\dagger_f \psi_f) a_+  \nonumber \\
{\rm{(scaling~dimension~5/2).~~~~~~~~ ~~~~~~~}}
\eea
We see that the leading irrelevant operator involves both spin and flavor degrees of freedom and has a total scaling dimension of 5/2. It leads to a conductance scaling  $\delta G \sim T^{\frac{3}{2}}$, as observed in Fig.~\ref{graph_table1}(c,f,i) (see graphs for unperturbed cases).

\section{Perturbations}
\label{se:perturbations}
Having analyzed the effective Hamiltonian and the resulting physical properties at the SO(5) fixed point, as summarized in the first column  of Table.~\ref{Tb:1}, we now discuss perturbations summarized in the remaining columns  of Table.~\ref{Tb:1}.
\subsection{Magnetic field} 
\label{se:magneticdf}
\begin{figure*}[t]
	\centering
	\includegraphics[scale=0.22]{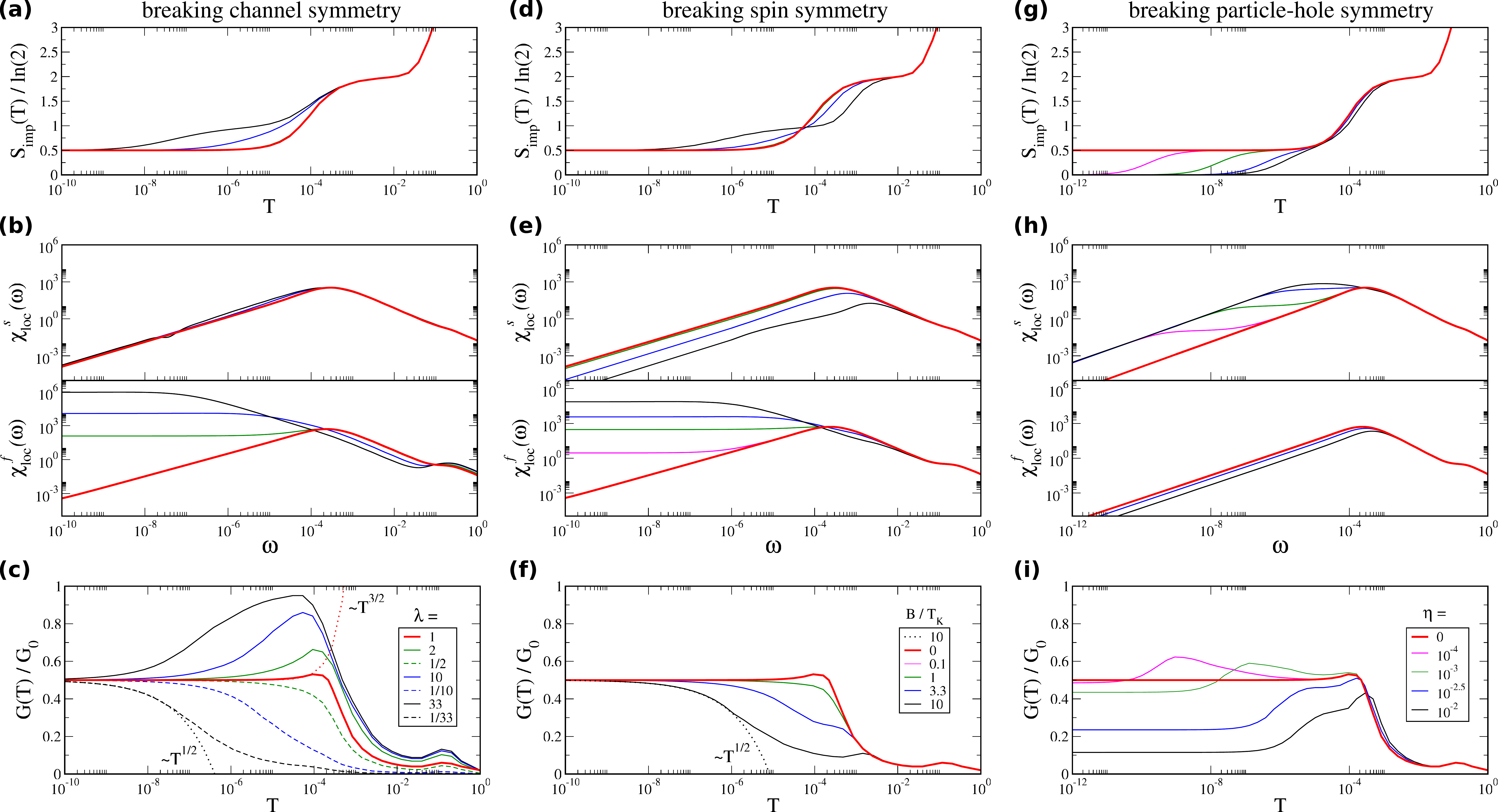}
	\caption{NRG results for different perturbations and different quantities. Different columns stand for different perturbations: column 1- channel asymmetry, column 2-magnetic field $B \ne 0$,  column 3- particle hole symmetry breaking $\eta \ne 0$. Different rows stand for different observables: row 1- the impurity's entropy, row 2-the spin and flavor susceptibilities, and row 3-conductance. The LR asymmetry $\lambda$ is defined as $\lambda=\xi^2 t_L^2/t_B^2$ where $\xi$ is found numerically such that at $t_B=\xi t_L$ the system displays LR symmetry; we vary $\lambda$ for fixed $t_B^2(1+\lambda)=0.02$. In our NRG calculations we measure energy in units of the band width $D=1$, set $E_c=0.1$; for further details see Ref.~\onlinecite{mitchell2020so}.}
	\label{graph_table1}
\end{figure*}

In the middle column of Fig.~\ref{graph_table1} we show NRG results for the entropy, susceptibilities and conductance in the presence of a magnetic field $B$. We can see that a residual $\frac{1}{2} \log 2$ entropy persists as $T \to 0$, implying a NFL state. We also observe a $\log 2$ intermediate temperature plateau, indicating the quenching of the spin degeneracy at $T < B$, followed by a NFL behavior. While neither the $\frac{1}{2} \log 2$ entropy at $T \to 0$ nor the  $\omega$-scaling of the spin-susceptibility are affected by $B$, the flavor susceptibility $\chi_f$ is no longer linear in $\omega$, and becomes a constant instead.

These behaviors can be accounted for within the Majorana fermion framework. The magnetic field term $-i B d_{+} d_{-}$ can be  incorporated into the fixed point Hamiltonian \eqref{Ekhamiltonian1new}. It can be readily combined with the $-2i Q_\perp^\perp d_+ a_-$ term by a  Majorana rotation, \bea
\label{eq:rotB}
\begin{pmatrix}
	a_-'\\
	d_-'
\end{pmatrix}=
\begin{pmatrix}
	\cos \alpha_B   & \sin \alpha_B\\
	-\sin \alpha_B  & \cos \alpha_B
\end{pmatrix}
\begin{pmatrix}
	a_-\\
	d_-
\end{pmatrix},
\eea
where $\sin \alpha_B=\frac{B}{\sqrt{(2Q_{\perp}^{\perp})^{2}+B^{2}}}$, yielding
\bea
\mathcal{H}^*_{SO(5)}-iB d_{+} d_{-}&=&H_0+i J_\perp'(\cos(\alpha_B) d_-'+\sin(\alpha_B) a_-') \chi_+\nonumber \\
&-& i 2Q' d_+ a_-',
\eea
where $Q'=\sqrt{(Q_{\perp}^{\perp})^{2}+(B/2)^{2}}$.
We first notice that $a_{+}$ remains  free (see scheme in column 2 of Table.~\ref{Tb:1}), supporting the observed $\frac{1}{2}\log{2}$ entropy at $T \to 0$. Although the Majorana rotation Eq.~(\ref{eq:rotB}) maintains  the system in a NFL state, it does modify certain correlation functions. 

The operator expansion of the impurity spin operator $S^z$, to leading order in $B$, remains as in Eq.~(\ref{eq:Szsf}), governed by the dimension-1 operator $\psi^\dagger_s \psi_s$, so that the spin susceptibility remains FL-like. 

On the other hand, the flavor impurity operator $T^z=i a_- a_+$ now obtains a diagonal component in terms of the rotated projectors $\mathcal{P}_{0,1}$, which are defined as in Eq.~(\ref{projectors}), but in terms of the rotated Majorana pair $(d_+,a_-')$ instead of $(d_+,a_-)$.  In the expression $T^z=i a_- a_+$, the Majorana $a_+$ is free, and $a_- = \cos \alpha_B a_-'-\sin \alpha_B d_{-}'$, contains a component $\propto \sin \alpha_B \propto B/(2Q_\perp^\perp)$ which is diagonal in the low and high energy subspaces associated with the $Q'$ term. Thus,
\bea
(T^z_{(+B)})_{00} \sim -i \frac{B}{2Q_\perp^\perp \sqrt{T_K}} \chi_+ a_+
\nonumber \\
~~~~~~~~{\rm{(scaling~dimension~1/2). ~~~~~~~}}
\eea
Hence, the magnetic field causes the  flavor operator to scale like the fermion field $\chi_+$, and acquire a scaling dimension $1/2$. As a result $\chi_f(t) \sim \frac{1}{t}$ ($\chi_f(\omega) \sim {\rm{const}}$). 

To address the effect of the magnetic field on the conductance, we identify the leading irrelevant operator. In the absence of $B$ the leading operator of dimension 5/2 in Eq.~(\ref{eq:leading53}) resulted from combining  $\delta H_s$ and $\delta H_f$, each of which individually bridges the low- and high-energy sectors of $Q_\perp^\perp$. However, in the presence of $B$ and the associated Majorana rotation, $\delta H_f$ alone contributes to the leading irrelevant operator
\be
\delta H_{irr}^{(+B)}=(\delta H_f)_{00} = -i (V_z-v_F) \frac{B}{2 Q_\perp^\perp \sqrt{T_K}} (\chi_+ \psi^\dagger_f \psi_f) a_+,
\ee
which has scaling dimension $\Delta=3/2$. Thus, the conductance scaling is modified into a $\delta G \sim T^{1/2}$ form. This is indeed observed in Fig.~\ref{graph_table1}(f).

\subsection{Breaking channel symmetry}
\label{channel}
We now consider  broken LR symmetry while still preserving PH symmetry. Among the list of operators in Table.~\ref{Table:sw}, this case includes $J_-$ and $V_z^-$. The most relevant term  is $\delta H_- =- i J_- d_+ \chi_-$,  see Eq.~\eqref{Ekhamiltonian2}. This extra coupling between $\chi_-$ and $d_+$, shown in Table.~\ref{Tb:1} column 3, still leaves the $a_+$ Majorana decoupled, implying a fixed point with $\frac{1}{2}\log 2$ entropy.

To determine the low energy behavior of the spin susceptibility, we look at the operator expansion of $S^z$, which in the absence of $J_-$, is given by the dimension-1 operator $\psi^\dagger_s \psi_s$ in Eq.~(\ref{eq:Szsf}). The leading order expansion of $S^z=i d_- d_+$ is obtained from Eq.~(\ref{eq:pert_theo_op}) where $H_{10},H_{01} \propto J_-$. However, due to anti-commutation relations, \eqref{eq:pert_theo_op} yields a vanishing contribution in this case. Hence the FL-like behavior of $\chi_s$ persists.

As for the flavor susceptibility, the flavor impurity operator $T^z=i a_- a_+$, now combines with $J_-$ according to Eq.~(\ref{eq:pert_theo_op}), to yield a dimension-1/2 operator of the form 
\be
\label{TzJm}
T^z \sim -i \frac{J_-}{2Q_\perp^\perp} \chi_- a_+.
\ee
Namely, the flavor susceptibility acquires NFL behavior at low energy.

Finally, the leading irrelevant operator stems from $\delta H_f$ combined with $H_-$ in Eq.~(\ref{eq:pert_theo_Ham}). It yields a dimension $3/2$ operator 
\be
\label{hirrJm}
\delta H_{irr} \sim -i\frac{J_-(V_z-v_F)}{2Q_\perp^\perp} \chi_-(\psi^\dagger_f \psi_f)a_+,
\ee
which leads to a $T^{1/2}$ scaling of the conductance as confirmed in Fig.~\ref{graph_table1}(c).

These low energy limits of the entropy, spin- and flavor-susceptibilities, and conductance, are summarized in Table.~\ref{Tb:1}, column 2. 

We now further discuss these results from our NRG simulations. In Fig.~\ref{graph_table1}(a)  we see the impurity entropy as a function of temperature for $(n_g,\eta)=(\frac{1}{2},0)$ and for various ratios between the left/right tunnelings  $\lambda \propto t_{L}^2/t_{B}^2$. While the $T=0$ fixed point entropy remains $\frac{1}{2}\log 2$ in all cases, channel asymmetry changes the way in which the entropy goes to its fixed point value. As asymmetry grows, two  successive crossovers appear: one from $\log(4)$ to $\log(2)$ on a scale $T_K^1$, and a second from $\log(2)$ to $\log(2)/2$ at a scale $T_K^2<T_K^1$. This is in contrast to the symmetric case which shows a single drop from $\log(4)$ to $\log(2)/2$. In the strongly asymmetric limit one can associate~\cite{zarand2006quantum,mitchell2012two} the first drop with a 1CK screening of the spin degree of freedom; while the $\frac{1}{2}\log 2$ drop is a nontrivial signature of a 2CK partial screening and relates to the flavor degree of freedom.
This claim is supported by our NRG calculation of the magnetic susceptibility in Fig.~\ref{graph_table1}(b), where only one crossover is observed at $T_K^{1}$. This indicates that the second crossover at $T_K^2$ for large asymmetry occurs in the flavor sector.

\subsection{Breaking Particle-hole symmetry}

PH symmetry is broken away from the point $(n_{g},\eta)=(\frac{1}{2},0)$. As can be seen in our NRG calculations in the right column of  Fig.~\ref{graph_table1}, in this case the NFL is destabilized, with a drop of the entropy to zero in Fig.~\ref{graph_table1}(g).   Notice that the flavor field which is turned on for $n_g \ne 0$, breaks PH symmetry as well as LR symmetry. Thus we should consider all the corresponding terms in Table.~\ref{Table:sw} $(V_\perp,Q_z;\phi;J_-,Q_z^-,V_z^-,\phi^-; B_f)$. We  focus on a subset of these perturbations which leads to a relevant instability of the SO(5) NFL state.

As in Sec.~\ref{se:magneticdf}, we begin with incorporating the Majorana-Majorana flavor coupling, $i B_f a_- a_+$, into the fixed point Hamiltonian. It can be combined with the $-2 i Q_\perp^\perp d_+ a_-$ term by another
Majorana rotation, \bea
\label{eq:rotB2}
\begin{pmatrix}
	\tilde{d}_+\\
	\tilde{a}_+
\end{pmatrix}=
\begin{pmatrix}
	\cos \alpha_{B_f}   & \sin \alpha_{B_f}\\
	-\sin \alpha_{B_f}  & \cos \alpha_{B_f}
\end{pmatrix}
\begin{pmatrix}
	d_+\\
	a_+
\end{pmatrix},
\eea
where $\sin \alpha_{B_f}=\frac{B_f}{\sqrt{(2Q_{\perp}^{\perp})^{2}+B_f^{2}}}$, yielding
\bea
\mathcal{H}^*_{SO(5)}-iB_f a_{+} a_{-}&=&H_0+i J_\perp d_- \chi_+\nonumber \\
&-& i 2 \tilde{Q} \tilde{d}_+ \tilde{a}_-,
\eea
where $\tilde{Q} =\sqrt{(Q_{\perp}^{\perp})^{2}+(B_f/2)^{2}}$.
We first notice that instead of $a_{+}$, at this stage $\tilde{a}_{+}$ is free (see Majorana scheme in column 4 of Table.~\ref{Tb:1}). However,  adding generic perturbations that break PH and LR symmetry, no Majorana remains free. We identify three perturbations, $V_\perp, Q_z$ and $J_-$, that when turned on, take the form of a  dimension-1/2 coupling  $i \tilde{a}_{+} \chi_-$. Explicitly, keeping only relevant operators in  $H_{V_\perp,Q_z}+\delta H_-$ in Eqs.~(\ref{Ekhamiltonian2}) and (\ref{Ekhamiltonian3}), we find
\bea
\label{eq:VV}
H_{V_\perp,Q_z}+\delta H_- & \to & -i \mathcal{V} \tilde{a}_{+} \chi_-, \nonumber \\
\mathcal{V} &=& \cos  \alpha_{B_f} (V_\perp+Q_z) -\sin  \alpha_{B_f}  J_- .
\eea
Due to the coupling $\mathcal{V}$ which is generically finite when PH symmetry is broken,  the so-far-free Majorana $\tilde{a}_+$ is coupled to the conduction electrons, hence the residual entropy is quenched, and the system turns into a FL, see Fig.~\ref{graph_table1}(g). In this case, resorting to the methods developed above, one can show that the system shows no anomalous NFL behavior, i.e., the impurity spin and flavor operators aquire scaling dimension-1, and the leading irrelevant operator has scaling dimension 2. This is summarized in column 4 of Table.~\ref{Tb:1}. However, as we discuss next, the NFL state can be recovered even in the PH and LR broken phase, upon fine tuning.

\section{Emergence of NFL line from SO(5) point in the $(n_g-\eta)$ phase diagram}
The coupling constant $\mathcal{V}$ in Eq.~(\ref{eq:VV}) depends on the tunneling amplitudes of the lead-dot-box model through the explicit forms of the various couplings, obtained through the SW transformation in Appendix \ref{App:SW}. The condition
\be
\label{eq:V0}
\mathcal{V}(J_-,V_\perp,Q_z,B_f)=0,~~~(\rm{NFL})
\ee
implies that the system is fine tuned to  a NFL state with a residual $\frac{1}{2} \log 2$ entropy. 

\begin{figure}[t]
	\centering
	\includegraphics[scale=0.47]{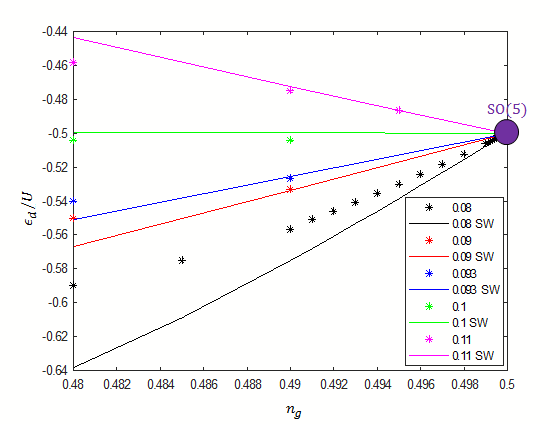}
	\caption{Numerical solutions of Eq.~(\ref{eq:V0}) qualitatively fitted to NRG results, for $t_{L}^{NRG}=0.08,0.09,0.1,0.11,0.092235$ and $t_{R}^{NRG}=0.12$, in the proximity of the $n_{g}=\frac{1}{2}$ point.
	}
	\label{gf:figfit}
\end{figure}

For a given  $t_{L}/t_{B}$ one can solve the equation $\mathcal{V}(\epsilon_d,U,E_c,t_L, t_B)=0$ for $n_{g}$ as a function of $\eta$. The resulting function is a NFL curve in the $(n_{g},\eta)$ phase diagram which approaches the SO(5) point. 
This is plotted in Fig.~\ref{gf:figfit}, where we solved Eq.~(\ref{eq:V0}), $\mathcal{V}(\epsilon_d,U,E_c,t_L^{NRG} \alpha, t_B^{NRG} \alpha)=0$ for different tunneling ratios. The cutoff scheme in NRG is different, allowing one fitting parameter $\alpha$ which is found to give optimal fitting for $\alpha \cong 0.11$.  We see that these curves qualitatively match the observed NRG NFL lines in the proximity of the $n_{g}=0.5,\eta=0$ point. While our  NRG results of Ref.~\onlinecite{mitchell2020so} extend in the entire ($n_{g},\eta$) phase diagram, which is periodic in $n_g \to n_g+1$ (see Fig.~1 in Ref.~\onlinecite{mitchell2020so}), our SW mapping which considered a limited number of charge states is restricted to the vicinity of point $n_g=1/2$.

While the fit shows the various NFL curves emanating from the SO(5) point with a LR-asymmetry dependent slope, the Toulouse limit approach is expected only to reproduce reliably the scaling properties (summarized in Table.~\ref{Tb:1})), but not the detailed dependence on model parameters.

Along the NFL curve, the Majorana $\tilde{a}_{+}$ remains decoupled. At the SO(5) point, $B_f=0$, it coincides with $a_+$ from the flavor sector [see Eq.~(\ref{eq:rotB2})]. As $B_f$ increases, the free Majorana tends towards $d_+$ from the spin sector. This indicates a smooth rotation in the spin and flavor  space~\cite{lebanon2003enhancement,lebanon2003coulomb,anders2004coulomb}.

\subsection{Physical properties}
We see that if we deviate from the SO(5) point along a specific direction, the system's NFL nature persists. We then have a free Majorana fermion, $\tilde{a}_{+}$, as illustrated in the scheme in column 5 of Table.~\ref{Tb:1}). Rather than the SO(5) NFL, we have generically on the NFL line a behavior similar to the spin-2CK NFL state, as implied by the physical signatures  discussed next. 

Using Eqs.~(\ref{replacement}) and (\ref{eq:rotB2}), the spin operator contains the terms $S^{z}=id_{-}d_{+}=i\frac{\chi_{+}}{\sqrt{T_{K}}}(\cos(\alpha_{B_f})\tilde{d}_{+}+\sin(\alpha_{B_f}) \tilde{a}_{+})$. Keeping only its low energy component, 
\be
(S^z)_{00} \sim \frac{1}{\sqrt{T_K}} \sin (\alpha_{B_f}) i \chi_{+} \tilde{a}_+ .
\ee
Comparing with Eq.~(\ref{eq:spin-2CK}) in the spin-2CK case, the spin susceptibility along the NFL curve contains a NFL component $\propto  \sin^2 (\alpha_{B_f})$ which vanishes at the SO(5) point, $\chi_s(\omega) \propto {\rm{const}} \times \sin^2 (\alpha_{B_f})+\mathcal{O}(\omega)$.

As for the flavor susceptibility, since we generically have a finite $J_-$ operator on the NFL line, we have the same dimension-1/2 operator as in Eq.~(\ref{TzJm}) leading to a NFL flavor susceptibility.  Similarly, there is a dimension 3/2 leading irrelevant operator $\propto J_-$ as in Eq.~(\ref{hirrJm}).

\section{Summary}
We revisited a quantum impurity problem describing a quantum dot device where two channels of electrons interact both with an impurity spin and with an additional "flavor" impurity. Using bosonization and refermionization, we constructed a consistent picture describing the coupling of the impurity degrees of freedom with the conduction electrons, and allowing to compute the various observables. Our field theory results are consistent with our numerical renormalization group calculations, and with a novel non-Fermi liquid fixed point exhibiting SO(5) emergent symmetry.

So far, it was believed that the device displays exotic two-channel Kondo behavior upon tunning the channel asymmetry to zero. Our study shows that  the NFL behavior is much more robust and extends to generic values of the channel asymmetry, upon tuning of the quantum dot level position. As demonstrated here in great detail, we reached this understanding from the high symmetry fixed point. Using NRG, as already demonstrated in Ref.~\onlinecite{mitchell2020so}, we found how this SO(5) fixed point connects in the phase diagram to the more conventional spin-two-channel Kondo state.
Our predictions within the systems's complex phase diagram can be tested experimentally, in terms of conductance~\cite{potok2007observation,keller2015universal}, and also more recent entropy measurements~\cite{hartman2018direct,sela2019detecting,kleeorin2019measuring}.

\section{Acknowledgements}
AKM and ES acknowledge the Stewart Blusson Quantum Matter Institute (UBC) for travel support. AKM acknowledges funding from the Irish Research Council Laureate Awards 2017/2018 through grant IRCLA/2017/169.
IA acknowledges financial
support from NSERC Canada Discovery Grant 0433-2016. ES acknowledges support from ARO (W911NF-20-1-0013), the Israel Science Foundation grant number 154/19 and US-Israel Binational Science Foundation (Grant No. 2016255).


%

\appendix
\section{Schrieffer-Wolff (SW) transformation}
\label{App:SW}
\begin{figure}
	\includegraphics[width=9cm]{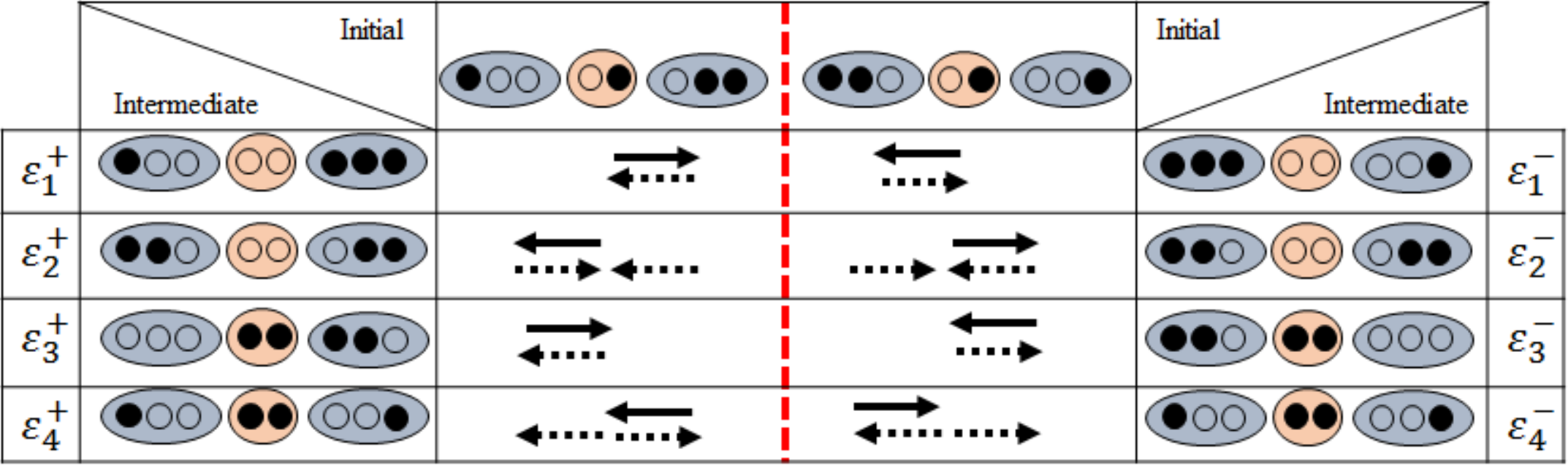}
	\caption{Schematic representation of  virtual transitions considered in our SW expansion. The two middle columns of the first row represent the two possible initial charge states $(N_{0},N_{0}+1)$ on the left and $(N_{0}+1,N_{0})$ on the right. The red circle represents the dot, the blue ellipse represents the box/leads, and the small empty or full circles indicate occupancy by electrons. The two outer columns show the  intermediate states. Full arrows  represent the direction in which electrons hop to reach the intermediate state, and dashed arrows represent the electronic hopping yielding  final state. The intermediate states energy is denoted by $\varepsilon_{i}^{\pm}$. For instance, the process that begins with an initial state of $(N_{0}+1,N_{0})$ becomes $(N_{0}+1,N_{0}-1)$ after an electron hops from the box to the dot, through the intermediate state of  energy $\varepsilon_{3}^{-}$.}
	\label{App:process}
\end{figure}

We derive the effective Hamiltonian \eqref{hamiltonianlehur} among the various additional interaction terms summarized in Table.~\ref{Table:sw}, starting from the lead-dot-box model. 
The system is initially either in the $(N_0,N_0+1)$ or $(N_0+1,N_0)$ charge states of the lead and box, with one electron (with spin up or down) in the small dot. To second order in the tunneling amplitude, depending on the initial state, the system visits one of the four intermediate states depicted in Fig.~\ref{App:process}. As long as the lead and box are equally treated as left and right reservoirs, the left and right sides of Fig.~\ref{App:process} are related by left-right transformation. Here, full-line arrow and dashed arrow represent the first and second tunneling event. For later use we detail the associated energies at zero tunneling. The energy is obtained from Eqs.~(2),~(3), $E=E_c(N_B-N_0-n_g)^2+\epsilon_d n_d+U n_\uparrow n_\downarrow$. For the two initial states $E^{(N_0,N_0+1)}=E_c(1-n_g)^2+\epsilon_d$ and $E^{(N_0+1,N_0)}=E_c n_g^2+\epsilon_d$. We assume without loss of generality that $n_g \le 1/2$, so the ground state energy is $E_{GS}=E^{(N_0+1,N_0)}$. For example, the energy of the first intermediate state in Fig.~\ref{App:process}, in which an electron tunnels from the small dot to the box, is $\varepsilon_{1}^+=E_c(2-n_g)^2$, while $\varepsilon_{1}^-=E_c n_g^2$. We define $E_{j}^{\pm} \equiv E_{GS}- \varepsilon_{j}^\pm$ $(j=1,2,3,4)$, which are given explicitly by 
\bea
\label{e1234}
E_{1}^{+}&=&\epsilon_{d}+4E_{c}(n_{g}-1),\\
E_{2}^{+}&=&\epsilon_{d}+E_{c}(2n_{g}-1)=E_{2}^{-},\nonumber\\
E_{3}^{+}&=&-\epsilon_{d}+E_{c}(2n_{g}-1)-U,\nonumber\\
E_{4}^{+}&=&-\epsilon_{d}-U=E_{4}^{-},\nonumber\\
E_{1}^{-}&=&\epsilon_{d},\nonumber\\
E_{3}^{-}&=&-\epsilon_{d}-E_{c}(2n_{g}+1)-U.\nonumber
\eea

To perform perturbation theory in the hybridization, we start from the ALS model, 
\bea
\label{eq:HALSNEW}
H_{ALS}&=& \sum_{\alpha = L,R} \sum_{k, \sigma} \epsilon_{k} \psi^\dagger_{ k \alpha \sigma}\psi_{k \alpha \sigma}+  E_c ( T^z_{B}- N_0-n_{g})^2 \nonumber  \\ 
&+&H_{\rm d}+\sum_{k\alpha\sigma} t_{\alpha} (d_\sigma^\dagger \psi_{k\alpha\sigma} T^-_\alpha  +H.c.),
\eea
where in comparison to Eq.~(\ref{eq:HALS1}), we added another pseudo-spin $T$-operator to the lead for convenience. Denoting the lead as the left ($L$) reservoir, and the box as the right ($R$) reservoir, we thus have two pseudo-spin operators $T_{L,R}$, whose $z$ component measures whether the corresponding reservoir is in the $N_0+1$ state ($T_\alpha=1/2$) or $N_0$ state ($T_\alpha=-1/2$) and $T^+$ increases the charge from $N_0$ to $N_0+1$. 

Treating the tunneling operators as a perturbation, $H_{ALS}=\mathcal{H}_{0}+H_{hyb}$ where $\mathcal{H}_{0} =H_0+H_B+H_d$, to second order the effective Hamiltonian is
\be
\delta H_{eff}=H_{hyb}\frac{1}{E-\mathcal{H}_{0}}   H_{hyb},
\ee
where $E=E_{GS}$,
giving
\bea
\delta H_{eff}=&&\sum_{k,k',\alpha,\beta,\sigma,\sigma'}t_{\alpha}t_{\beta}(d^{\dagger}_{\sigma}c_{k\alpha \sigma}T^{-}_{\alpha} \frac{1}{E-\mathcal{H}_{0}} T^{+}_{\beta}c_{k'\beta\sigma'}^{\dagger}d_{\sigma'}\nonumber\\
&&+T^{+}_{\alpha}c_{k\alpha\sigma}^{\dagger}d_{\sigma}\frac{1}{E-\mathcal{H}_{0}}d^{\dagger}_{\sigma'}c_{k'\beta\sigma'}T_{\beta}^{-}).
\eea
The first term corresponds to a hole processes with an empty small dot intermediate state (states $\varepsilon_{1,2}^\pm$ in Fig.~\ref{App:process}) while the second corresponds to particle processes involving a doubly occupied small dot (states $\varepsilon_{3,4}^\pm$ in Fig.~\ref{App:process}). The energy denominator represents (minus) the energy difference between the virtual and initial state as given in Eq.~(\ref{e1234}). 
We now reduce the two pseudo-spin operators back to a single one representing the pair of states $(N_0,N_0+1)$ or $(N_0+1,N_0)$, defined as
\bea
T^{+}=T_{L}^{-}T^{+}_{R}, \ \ T^{-}=T^{+}_{L}T^{-}_{R}, \ \ T^{z}=T_{R}^z=-T_{L}^z.
\eea
We also define  projection operators, $P_{+},P_{-}$, into the $(N_0,N_{0}+1)$, $(N_0+1,N_{0})$ states, respectively,
\bea
P_{+}=1/2+T^{z},~~~~~ \ \ P_{-}=1/2-T^{z}.
\eea
The effective interactions can be separated as $\delta H_{eff}=H_++H_-$ depending on whether the system starts in the $+$ or $-$ state, each of which contains 4 terms linked with the processes in Fig.~\ref{App:process}. For the $(N_0,N_0+1)$ initial state we have
\bea
H_{+}&=&[\sum_{k,k',\sigma,\sigma'}t_{R}^{2}\frac{1}{E_{1}^{+}}(d^{\dagger}_{\sigma}c_{k R\sigma}T^{-}_{R}T^{+}_{R}c_{k' R \sigma'}^{\dagger}d_{\sigma'})\\
&+&\sum_{k,k',\alpha,\sigma,\sigma'}t_{L}t_{\alpha}\frac{1}{E_{2}^{+}}(d^{\dagger}_{\sigma}c_{k\alpha \sigma}T^{-}_{\alpha}T^{+}_{L}c_{L,k',\sigma'}^{\dagger}d_{\sigma'})\nonumber\\
&+&\sum_{k,k',\sigma,\sigma'}t_{L}^{2}\frac{1}{E_{3}^{+}}(T^{+}_{L}c_{k L\sigma}^{\dagger}d_{\sigma}d^{\dagger}_{\sigma'}c_{k' L\sigma'}T_{L}^{-})\nonumber\\
&+&\sum_{k,k',\alpha,\sigma,\sigma'}\frac{t_{\alpha}t_{R}}{E_{4}^{+}}(T^{+}_{\alpha}c_{k\alpha\sigma}^{\dagger}d_{\sigma}d^{\dagger}_{\sigma'}c_{k' R\sigma'}T_{R}^{-})]P_{+}.\nonumber
\eea
Similarly, from the $(N_{0}+1,N_{0})$ state (right side in Fig.~\ref{App:process}) we generate the interactions
\bea
H_{-}&=&[\sum_{k,k',\sigma,\sigma'}t_{L}^{2}\frac{1}{E_{1}^{-}}(d^{\dagger}_{\sigma}c_{ k L\sigma}T^{-}_{L}T^{+}_{L}c_{k' L\sigma'}^{\dagger}d_{\sigma'})\\
&+&\sum_{k,k',\alpha,\sigma,\sigma'}t_{R}t_{\alpha}\frac{1}{E_{2}^{-}}(d^{\dagger}_{\sigma}c_{k \alpha \sigma}T^{-}_{\alpha}T^{+}_{R}c_{k' R\sigma'}^{\dagger}d_{\sigma'})\nonumber\\
&+&\sum_{k,k',\sigma,\sigma'}t_{R}^{2}\frac{1}{E_{3}^{-}}(T^{+}_{R}c_{k R\sigma}^{\dagger}d_{\sigma}d^{\dagger}_{\sigma'}c_{k' R\sigma'}T_{R}^{-})\nonumber\\
&+&\sum_{\alpha,k,k',\sigma,\sigma'}\frac{t_{\alpha}t_{L}}{E_{4}^{-}}(T^{+}_{\alpha}c_{k\alpha\sigma}^{\dagger}d_{\sigma}d^{\dagger}_{\sigma'}c_{k' L\sigma'}T_{L}^{-})]P_{-}.\nonumber
\eea
Defining local fermion operators $\psi_{\alpha \sigma}=\sum_k c_{k\alpha\sigma}$ and impurity spin $\vec{S}=d^\dagger_\sigma \frac{\vec{\sigma}_{\sigma \sigma'}}{2} d_{\sigma'}$, the resulting effective Hamiltonian can be written as $\delta H_{eff}=\sum_{i} \lambda_i \mathcal{O}_i$, where the  local operators $\mathcal{O}_i$ and coupling constants $\lambda_i$ are listed in Table.~\ref{Table:sw}. Each operator has a well defined left-right (LR) transformation and PH symmetry. 

We separated the coupling constants into (i) a pair of operators that transmit charge from left to right ($V_\perp$ and $Q_\perp$) containing $T^\pm$, (ii) 4 operators which are LR even ($J,V_z,Q_z,\phi$), (iii) LR odd versions of the former four ($J_-,V_{z-},Q_{z-},\phi_-$), and (iv) the flavor field $T^z$.

Following a tedious but straightforward algebra, the coupling constants can be obtained.
The explicit form of the pseudospin-flip operators are
\bea
\label{fullhamiltonian3perp}
Q_{\perp}&=&-t_{R}t_{L}\left(\frac{1}{E_{2}^{-}}+\frac{1}{E_{4}^{-}}\right),\nonumber\\
V_{\perp}&=&-t_{R}t_{L}\left(\frac{1}{E_{2}^{+}}-\frac{1}{E_{4}^{+}}\right).
\eea
The four couplings which have a LR odd version, such as $J$, can be written as $J=J^L+J^R$ and $J_-=J^L-J^R$, and similarly for $V_z,Q_z$ and $\phi$ where
\bea
\label{fullhamiltonian3}
J^{R}&=&-\frac{t_R^2}{2} \left(\frac{1}{E_{1}^{+}}+\frac{1}{E_{2}^{-}}+\frac{1}{E_{3}^{-}}+\frac{1}{E_{4}^{+}} \right),\nonumber\\
Q_{z}^{R}&=&\frac{t_R^2}{2}\left( (\frac{1}{E_{1}^{+}}-\frac{1}{E_{2}^{-}})-(\frac{1}{E_{3}^{-}}-\frac{1}{E_{4}^{+}})\right), \nonumber\\
\phi^{R}&=&-\frac{t_R^2}{8}\left((\frac{1}{E_{1}^{+}}+\frac{1}{E_{2}^{-}})-(\frac{1}{E_{3}^{-}}+\frac{1}{E_{4}^{+}})\right),\nonumber\\
V_{z}^{R}&=&\frac{t_R^2}{4}\left((\frac{1}{E_{3}^{-}}-\frac{1}{E_{4}^{+}})+(\frac{1}{E_{1}^{+}}-\frac{1}{E_{2}^{-}})\right),\nonumber
\eea
and $J^L,Q_z^L,\phi^L,V_z^L$ are obtained from these expressions by interchanging $E_j^+ \leftrightarrow E_j^-$ and $t_R \leftrightarrow t_L$.  Similarly, the flavor field is given by $B^{R}=t_{R}^{2}(\frac{1}{E_{1}^{+}}-\frac{1}{E_{2}^{-}})$. This term correcting Eq.~(\ref{eq:Bf}) leads simply to a shift in the definition of $n_g$.

\section{PH transformation of bosonic and Majorana fields}
\label{App:ph}
Under the PH transformation Eq.~(\ref{ourPH}),  $\psi_{\alpha\uparrow} \to -i \psi^\dagger_{\alpha \downarrow}$ and $\psi_{\alpha\downarrow} \to i \psi^\dagger_{\alpha\uparrow}$. Rewriting this  in terms of boson fields and Klein factors using Eq.~(\ref{eq:psiphi}), we have
\bea
F_{\alpha\uparrow} e^{-i \phi_{\alpha\uparrow}} &\to& -i F^\dagger_{\alpha\downarrow} e^{i \phi_{\alpha\downarrow}}, \nonumber \\
F_{\alpha\downarrow} e^{-i \phi_{\alpha\downarrow}} &\to& i F^\dagger_{\alpha\uparrow} e^{i \phi_{\alpha\uparrow }}.
\eea
This transformation rule is consistent with
\be
F_{\alpha\uparrow} \to F^\dagger_{\alpha\downarrow},~~~~F_{\alpha\downarrow} \to + F^\dagger_{\alpha\uparrow},
\ee
and
\bea
\phi_{\alpha\uparrow} \to -\phi_{\alpha\downarrow}+\pi /2,~~~~
\phi_{\alpha\downarrow}\to -\phi_{\alpha\uparrow}-\pi /2.
\eea
Here we made a choice attaching  the minus sign in the PH transformation to the bosons rather than Klein factors. 

Moving to the $\mathcal{A}=c,s,f,sf$ bosons using Eq.~(\ref{eq:csfsf}), we have
\bea
\phi_{sf}  \to  \phi_{sf},~~~
\phi_s  \to  \phi_s + \pi,  \nonumber \\
\phi_f  \to  - \phi_f, ~~~
\phi_c  \to  - \phi_c.  
\eea

The relevant perturbation $V_\perp$ in Table.~\ref{Table:sw} is PH odd. It adds to the Hamiltonian the perturbation $-i V_\perp a_+ \chi_-$, which consists of the hermitian operator
\be
\epsilon = (a^\dagger+a)(\psi_{sf}-\psi_{sf}^\dagger),
\ee
where $a=F_f^\dagger T^-$. (The notation $\epsilon$ relates to our CFT solution of the problem~\cite{mitchell2020so}, where this field belonged to the Ising CFT.) We would like to show that $\epsilon$ is PH-odd. Explicitly
\be
\epsilon = (F_f T^-  + F_f^\dagger T^-)(F_{sf} e^{-i \phi_{sf}} + F_{sf}^\dagger e^{i \phi_{sf}}).
\ee
Now we perform the PH transformation on the pair of Klein factors,
\be
F_{sf}^\dagger F_f^\dagger = F^{\dagger}_{1\uparrow}  F_{2\uparrow} \to F_{1\downarrow}  F_{2\downarrow}^{\dagger} =-  F_{2\downarrow}^{\dagger} F_{1\downarrow }=-\mathcal{F}_{sf}^\dagger \mathcal{F}_f.
\ee
In the last equality we used the relation in Eq.~(\ref{zarand1998simple11}) and the unitarity of Klein factors. We conclude that under PH $F_f \to -F_f^\dagger$. Combining this with $T^+ \to T^-$, we  see that $\epsilon \to - \epsilon$.

\section{Weak coupling RG equations and flow towards the SO(5) fixed point}
\label{App:RG}
\subsection{Derivation of Eq.~(\ref{eq:ope})}
We begin with a derivation of Eq.~(\ref{eq:ope}) following the perturbative RG approach~\cite{cardy1996scaling}. Consider a fixed point theory described by a set of operators $\mathcal{O}_i$ having scaling dimensions $\Delta_i$, and satisfying the  operator product expansion (OPE)
\be
\mathcal{O}_i(x) \mathcal{O}_i(y) = \sum_k c_{ijk} \frac{ \mathcal{O}_k(y)}{(x-y)^{\Delta_i+\Delta_j-\Delta_k}}.
\ee
Now consider the fixed point Hamiltonian $\mathcal{H}^*$ perturbed by $\delta \mathcal{H} = \sum_i \lambda_i \mathcal{O}_i$. In our case all these operators are marginal, $\Delta_i=1$. Then the perturbative RG equations are determined by the OPE coefficients~\cite{cardy1996scaling}
\bea
d\lambda_i/d \ell = \sum_{jk}  c_{ijk} \lambda_j \lambda_k.  
\label{app:lambda}
\eea
We have 15 operators $\mathcal{O}_A = (\psi^\dagger T^A \psi) (f^\dagger T^A f)$ with $T_A$ given in Eq.~(\ref{Tmatrix}). PH symmetry reduces this to 10 operators. We can gain further insight into the properties of $\mathcal{O}_A$ by analyzing its two factors.
In Sec.~\ref{sec:su4vs} we defined $M^{ab}=f^\dagger T^{ab} f$ and $J^{ab}=\psi^\dagger T^{ab}\psi$, such that $A \in (ab), ~(a,b=1,...5, ~a<b)$. The impurity operators $M^{ab}$ satisfy $[T^{ab},T^{cd}]=-i (\delta_{bc} T^{ad}-\delta_{ac} T^{bd}-\delta_{bd} T^{ac}+\delta_{ad} T^{bc})$. The $J^{ab}$ operators satisfy a similar OPE. This can be seen if we define a vector of Majorana fields $\vec{\xi}$ via
\bea
\vec{\xi}&=&\\
&&(\chi_{s+},-\chi_{s-},\chi_{sf+},\chi_{f-},\chi_{f+},-\chi_{sf-},\chi_{c+},\chi_{c-}).\nonumber
\eea
By explicitly using the EK bosonization and refermionization one can show that $J^{ab}=i\xi_{a}\xi_{b}$ ($a,b=1,...,6$). The OPE of the $J^{ab}$'s then follows from Wick's theorem, and from this one can obtain the OPE satisfied by the $\mathcal{O}_A$ operators, leading to the RG equation Eq.~\ref{eq:ope}.

\subsubsection{Numerical integration}
\label{se:numericalRG}
\begin{figure}
	\includegraphics[width=7cm]{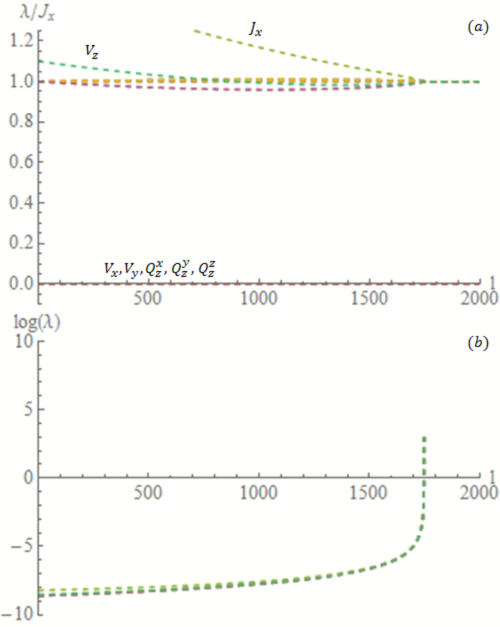}
	\caption{(a) the value of the couplings (there are 15 of them in total) normalized by $J_{x}$, as a function of the RG scale parameter $\ell$, for an initial value where all couplings are equal to $u=0.00018$, except for $J_{z}=1.5u$, $V_{z}=1.1u$ and $V_{x},V_{y},Q_{z}^{x},Q_{z}^{y},Q_{z}^{z}=0$. We see that all ratios of couplings flow to unity. (b) For the same RG flow we display on a logarithmic scale the absolute values of the couplings. We see that isotropy is reached together with strong coupling, in which the weak coupling RG equations become uncontrollable. } 
	\label{fig:rgflow}
\end{figure}

In Fig.~\ref{fig:rgflow} we integrated Eq.~(\ref{eq:ope}) starting with anisotropic finite values of the 10 PH symmetric couplings.  We see a flow to the  isotropic SO(5) fixed point. 

	\end{document}